\documentclass[aps,amsmath,amssymb,superscriptaddress,nofootinbib,twocolumn,floatfix]{revtex4-1}

\usepackage{graphicx}
\usepackage[usenames]{color}
\usepackage[colorlinks=true,linkcolor=blue,citecolor=blue,anchorcolor=green,urlcolor=blue]{hyperref}

\begin{document}

\title{
Local dynamics and thermal activation in the transverse-field Ising chain
}

\author{Jiahao Yang}
\affiliation{Tsung-Dao Lee Institute \& School of Physics and Astronomy, Shanghai Jiao Tong University, Shanghai 200240, China}

\author{Weishi Yuan}
\affiliation{%
 Department of Physics and Astronomy, McMaster University, Hamilton, Ontario, L8S 4M1, Canada
}%
\author{Takashi Imai}
\affiliation{%
 Department of Physics and Astronomy, McMaster University, Hamilton, Ontario, L8S 4M1, Canada
}%

\author{Qimiao Si}
\email{qmsi@rice.edu}
    \affiliation{Department of Physics \& Astronomy, Rice University, Houston, Texas 77005, USA}

\author{Jianda Wu}
\email{wujd@sjtu.edu.cn}
\affiliation{Tsung-Dao Lee Institute \& School of Physics and Astronomy, Shanghai Jiao Tong University, Shanghai 200240, China}

\author{M\'{a}rton Kormos}
\email{kormosmarton@gmail.com}
\affiliation{MTA-BME Quantum-Dynamics and Correlations Research Group,  E\"otv\"os Lor\'and Research Network (ELKH), Budapest University of Technology and Economics, 1111 Budapest, Budafoki \'{u}t 8, Hungary}

\begin{abstract}
There has been considerable recent progress in identifying candidate materials
for the  transverse-field Ising chain (TFIC), a paradigmatic model for quantum criticality.
Here, we study the local spin dynamical structure factor of different spin components
in the quantum disordered region of the TFIC.
We show that the
low-frequency local dynamics of the spins in the Ising- and transverse-field directions have strikingly
distinctive temperature dependencies. This leads to the thermal-activation gap for the
secular term of the
NMR $1/T_2^{\prime}$
 relaxation rate
to be half of that for the $1/T_1$ relaxation rate. Our findings reveal a new surprise in the nonzero-temperature dynamics
of the venerable TFIC model and
uncover a means to evince the material realization of the TFIC universality.
\end{abstract}

\date{\today}

\def\be{\begin{equation}}
\def\ee{\end{equation}}
\def\bea{\begin{eqnarray}}
\def\eea{\end{eqnarray}}
\def\erf{\eqref}
\newcommand{\expct}[1]{\left\langle #1 \right\rangle}
\newcommand{\expcts}[1]{\langle #1 \rangle}
\newcommand{\ud}          {\mathrm d}
\newcommand\eps           {\varepsilon}
\newcommand\sig           {\sigma}
\newcommand\w           {\omega}
\newcommand\fii           {\varphi}
\newcommand\mc            {\mathcal}
\newcommand\LL            {Lieb--Liniger }
\newcommand\p             {\partial}
\newcommand\kB             {k_\text{B}}
\newcommand\psid          {\psi^{\dagger}}
\renewcommand\th          {\theta}
\newcommand\kb            {\tilde k}
\newcommand\ob            {\tilde\omega}
\newcommand \rhop       {\rho^{\text{(r)}}}

\renewcommand{\vec}[1]   {|#1\rangle}
\newcommand{\cev}[1]   {\langle#1|}
\newcommand{\vev}[1]{\left\langle #1 \right\rangle}

\maketitle

{\it Introduction.~~}
While classical
matter freezes at zero temperature, quantum
many-body systems often display
multiple ground states due to the competition between different couplings.
Upon a continuous transformation from one ground state to another,
quantum criticality develops.
It is anchored by a quantum critical point (QCP) at zero temperature,
in contrast to a classical critical point that appears at
a thermally-induced phase transition.
Quantum criticality has emerged as a general organizing principle to understand
many of the richest phenomena that have been observed in quantum materials
\cite{special_2013,paschen_quantum_2021}.
These include the cuprates \cite{keimer_quantum_2015,Ramshaw_2015}, heavy fermion metals \cite{kirchner_colloquium_2020,coleman_quantum_2005,lohneysen_2007} and iron pnictides \cite{dai_iron_2009,licciardello_electrical_2019}.
One of the prominent features of quantum criticality is that it mixes spatial and temporal fluctuations.
While this intermixing complicates the description of quantum criticality, it also implies that
dynamical properties can be used to characterize the nature of quantum criticality.
As another outstanding feature of quantum criticality, 
approaching the QCP by a non-thermal control parameter ($g$)
and by temperature ($T$) represent two independent ways to examine its universal behavior.
As such, it is instructive to probe quantum criticality by analyzing dynamical properties as a function of temperature,
which can be conveniently studied experimentally.

A paradigmatic model for continuous quantum phase transitions is the transverse field Ising model in one spatial dimension
\cite{niemeijer_exact_1967,pfeuty_one-dimensional_1970,sachdev_2011,barouch_statistical_1971,suzuki_relationship_1971,suzuki_relationship_1976,kopp_criticality_2005}.
It represents a prototype setting to explore the properties of quantum criticality.
Yet, in spite of its venerable status, there is much about it that remains to be understood.
Suitable materials to study this model have only been emerging recently
\cite{coldea_quantum_2010,wang_tfic_quantum_2018,cui_tfic_quantum_2019,zhang_e8_observation_2020,zou_e_8_2021}.
They have allowed new experiments that are providing puzzling experimental results,
especially on dynamical properties at nonzero temperatures.
At the same time, calculating dynamical quantities near a QCP are always challenging.
That is also the case for the transverse field Ising chain,
notwithstanding the considerable efforts \cite{niemeijer_exact_1967,pfeuty_one-dimensional_1970,jianda_crossovers_2018}.
One of the particularly interesting quantities is the local dynamical structural factor \cite{jianda_E8_2014,unpublished,steinberg_nmr_2019},
which can be measured by the NMR longitudinal relaxation rate $1/T_1$ \cite{kinross_evolution_2014}.

In this work, we study the local dynamics that can be measured
by the NMR
 transverse relaxation rate $1/T_2$.
We consider the quantum disordered region at $g>g_c$, and calculate
$S^{xx}$ and $S^{zz}$ vs. $T$ at $\omega \ll {\kB }T /\hbar$
[for notations about $x$ and $z$, see Eq.\,\eqref{Hamiltonian} below.]
We find that they have
thermally-activated behavior with activation gaps
that differ by a factor of $2$.
This leads to the surprising and experimentally testable finding that
the activation gap for the secular term of $1/T_2$,
named $1/T_2^{\prime}$ (Ref.\,[\onlinecite{Jaccarino1965}]),
 will be half of that for $1/T_1$.

{\it The model.~~}
The Hamiltonian of the transverse field Ising chain is given by \cite{pfeuty_one-dimensional_1970}
\be
H_I  =  - J\sum_{i=1}^N\left( { {\sigma _i^z \sigma _{i + 1}^z }  + g  {\sigma _i^x } } \right)\,,
\label{Hamiltonian}
\ee
where $\sigma _i^x$ and $\sigma _i^z$ are Pauli matrices
associated with the spin components $S^\mu = \sigma^\mu/2, ({\mu=x,y,z})\;$on site $i$,
and $g$ is the coupling with the transverse field.
Below we shall refer to the $z$ (Ising) direction as {\it longitudinal} and to the $x$ direction as {\it transverse}.
At zero temperature, the system undergoes a quantum phase transition
when the transverse field is tuned across its QCP $g=g_c=1$.
The Hamiltonian can be conveniently converted to fermionic operators,
 $c_i$ and $c_i^\dag$ through Jordan-Wigner transformation.
After introducing a Bogoliubov rotation, as described in Appendix \ref{sec:J-W},
the Hamiltonian takes the canonical form $H_\text{I}=\sum_k \epsilon_k(\gamma_k^\dagger\gamma_k-\frac{1}{2})$
with single-particle energy dispersion
\be
\label{latdisprel}
\epsilon_k=2J\sqrt{1+g^2-2g\cos k}\,,
\ee
where momentum $k$ is dimensionless in all calculations.
At zero momentum, the gap $\Delta=2J|g-g_c|$ which vanishes at $g_\text{c}.$

In the vicinity of the QCP, the low energy effective description of the system is given by 
an Ising field theory obtained
as the scaling limit of the lattice Hamiltonian Eq.~\eqref{eq:H_gamma}.
In this limit the lattice spacing goes to zero, $a\to0,$ while $J\to\infty$ and $g\to1$ such that the energy gap
and the ``speed of light'' are kept fixed, $2J(1-g)=\Delta,$ $2Ja/\hbar=c.$
The resulting Hamiltonian describes a relativistic field theory of free Majorana fermions with mass $m=\Delta/c^2$
\be
\label{HFT}
{\mathcal H}_\text{I}=\int\ud x
\left[ \hbar c\frac{i}2(\psi\frac{\partial\psi}{\partial x}-\bar\psi \frac{\partial \bar \psi}{\partial x})
\pm \Delta\frac{i}2(\bar\psi\psi-\psi\bar\psi)\right]\,,
\ee
where the sign of second term is $+$ ($-$) for the paramagnetic (ferromagnetic) regime, corresponding to $g > g_c$ ($g < g_c$) in the lattice model.
{The field operators are related to the lattice operators as
$\stackrel{\scriptscriptstyle{\,(-)}}{\psi}\!\!\!\!\!(ja) = \frac1{\sqrt{2a}}\left(e^{\mp i\pi/4}c_j+e^{\pm i\pi/4}c_j^\dag\right).$}
The spectrum of the Ising field theory is built from multi-particle free fermion states with relativistic dispersion
for a single particle
\be
\varepsilon(p) = \sqrt{\Delta^2+p^2c^2} = \Delta \cosh\theta\,,
\ee
where {$p=\hbar k/a$}  and $\theta$ is the relativistic rapidity parameter.
The scaling limit of the transverse magnetization operator can be written as
\be
\label{epsilon}
\sigma_j^x = -2a\, \eps (x=ja) \equiv -2a\,i\,\bar\psi(x)\psi(x)\,.
\ee
The field theory operator $\sigma(x)$ corresponding to the order parameter is non-local with respect to the fermions
and cannot be expressed in a simple way in terms of the latter.
Assuming the standard field theory normalization, it is related to the lattice magnetization as ($\hbar=c=1$) \cite{pfeuty_one-dimensional_1970}
\be
\label{sbar}
\sigma(x=ja) = 2^{-1/24}e^{-1/8}\mathcal{A}^{3/2}a^{-1/8}\sigma^z_j=\bar s J^{1/8} \sigma^z_j\,,
\ee
where $\mathcal{A}=1.2824271291\dots$ is Glaisher's constant.

{\it Dynamic structure factor.~~}
We compute the
local spin dynamical
structure factor (DSF) for frequencies $\w\ll k_\text{B}T$
which are relevant to the NMR relaxation rates.
The DSF of the spin component $\alpha$ is given by
\begin{multline}
S^{\alpha\alpha}(\w,q)=\frac{-2}{1-e^{-\beta\w}} \mathrm{Im}\,\chi^{\alpha\alpha}(\w,q)=\\
\sum_l \int_{-\infty}^\infty\ud t\,e^{i\w t-iqla}
\langle \sig^\alpha_{l+1}(t) \sig^\alpha_1(0)\rangle_T\,,
\end{multline}
where $\beta=(\kB T)^{-1}$, and
$\chi^{\alpha\alpha}(\w,q)$ is the dynamical
spin susceptibility at the 
transferred energy
$\omega$ and momentum $q$.
In the field theory, we consider the continuum operators Eqs.~\eqref{epsilon}, \eqref{sbar}
and the summation over lattice sites is replaced by an integral over $x$.
The local DSF is
$S^{\alpha\alpha}(\w) = \int\frac{\ud q}{2\pi} S^{\alpha\alpha}(\w,q)$.
Both on the lattice and in the field theory,
we proceed with a Lehmann spectral representation.
Using the field theory language,
it reads
\be
S^{\alpha\alpha}(\w,q)=\frac1{\mc{Z}}\sum_{r,s=0}^\infty C^{\alpha\alpha}_{r,s}(\w,q)
\label{eq:series_text}
\ee
with $\alpha = x, y, z$ \cite{sigma_y} and
\begin{multline}
C^{\alpha\alpha}_{r,s}(\w,q)=\int\frac{\ud\th_1\dots\ud\th_r}{(2\pi)^rr!}\int\frac{\ud\th'_1\dots\ud\th'_s}{(2\pi)^ss!}e^{-\beta E_r}
(2\pi)^2\\
\delta(\w+E_r-E_s)\delta(q+P_r-P_s)
\,|\cev{\th_1\dots\th_r}\sig^\alpha(0,0)\vec{\th'_1\dots\th'_s}|^2\,,
\end{multline}
where the energy and momentum eigenvalues of the
multiparticle states $|\th_1,\dots,\th_n\rangle$
are
$E_n=\Delta\sum_i^n\cosh\th_i$ and $P_n=(\Delta/c)\sum_i^n\sinh\th_i$
($c$ is re-introduced here and following).
As described in Appendix \ref{sec:form-factor},
this series is a low-temperature expansion. It has term-by-term divergences that can be
regularized in a linked cluster expansion \cite{Essler_2009,gabor_2010},
${S}^{\alpha\alpha}(t,x)=\sum_{r=0,s=0}^\infty D^{\alpha\alpha}_{r,s}(t,x),$
where the finite terms $D^{\alpha\alpha}_{r,s}$ are certain linear combinations
of $C^{\alpha\alpha}_{r,s}$
and terms appearing in the expansion of the partition function.

{\it Local transverse DSF in the scaling limit.~~}
We now turn to truncated form factor series results of $S^{xx}(\omega)$
in the quantum disordered region with $\omega\ll \kB T\ll\Delta$.
The field theory operator corresponding to the transverse magnetization, $\eps=i\,\bar\psi(x)\psi(x)$ [cf. Eq.~\eqref{epsilon}],
is quadratic in the fermionic operators, so it only has nonzero form factors between states that
either have an equal number of particles or the particle number difference is 2. In particular,
\be
F_{2}^{\eps}(\th|\th')\equiv \langle\th|\eps(0)|\th'\rangle =
-i\frac{\Delta}{c} \cosh\left(\frac{\th-\th'}2\right)\,,
\ee
as can be obtained from the plane wave expansion of the fields [cf. Eq.~\eqref{psiexp}] in a straightforward way.

The first term in the series is the vacuum contribution $C_{00}^{xx} = |\langle0|\eps|0\rangle|^2\delta(\w)\delta(k)$,
i.e. 0 particle - 0 particle (0p - 0p) contribution.
In the field theory this is a divergent contribution
and requires renormalization (e.g. by normal ordering).
However, since we are interested in the finite (but small) $\w$ domain, we ignore this term.
The terms $D_{0,s}^{xx}$ and $D_{s,0}^{xx}$ contribute to frequencies $\w \ge s \Delta$ outside of our domain of interest,
 $0<\w\ll\Delta.$
 The first contributing term is thus
\begin{multline}
\label{C11}
C_{11}^{xx}(\w,q)=
\int\int\frac{\ud\th}{2\pi}\frac{\ud\th'}{2\pi}
|F_{2}^{\eps}(\th|\th')|^2
e^{-\beta \Delta\cosh\th}(2\pi)^2\\
\delta\left[q+\frac{\Delta}{c}(\sinh\th-\sinh\th')\right]\delta[\w+\Delta(\cosh\th-\cosh\th')]\,.
\end{multline}
Then after performing two integrals, the leading behavior of local transverse DSF
at $\omega\ll \kB T\ll\Delta$ region is (cf. Appendix~\ref{app:xxscaling})
\be
\label{C11asym}
C_{11}^{xx}(\w) \approx
 -\frac{\Delta}{\pi c^2} e^{-\frac{\Delta}{\kB T}}
 \left[\ln\left(\frac{\w}{4\kB T}\right)-\frac{\kB T}{2\Delta}+\gamma_\text{E}\right] \,,
\ee
where $\gamma_{E}$ is Euler's constant.
The main features of the result are the $\sim e^{-\Delta/\kB T}$ temperature dependence and the $\sim\ln(\w/\kB T)$ logarithmic frequency dependence.

We have computed $C_{11}^{xx}$, but since in $D_{11}^{xx}=C_{11}^{xx}-\mathcal{Z}_1C_{00}^{xx}$
the second term contributes to the static $\sim\delta(\w)$ response only,
the obtained result is equal to $D_{11}^{xx}$ for any nonzero $\w$.
The higher particle number contributions $D_{r,s}^{xx}$
with $\mathrm{max}(r,s)\ge2$ contain the factor $e^{-\beta E_r}\delta(E_r-E_s+\omega)$,
so they are exponentially suppressed at low temperature $\beta\Delta\gg1$
for frequencies $\w\ll\Delta$, in particular,
$D_{r,s}^{xx}\sim e^{-\mathrm{max}(r,s)\beta\Delta}$.
So we find that the leading order contribution to the local transverse DSF $S^{xx}(\omega)$
in the quantum disordered region with $\omega\ll \kB T\ll\Delta$ is given by Eq.~\erf{C11asym}.
An alternative derivation of the same result is given in Appendix \ref{app:alt}.


{\it Local transverse DSF in the spin chain.~~}
The truncated form factor expansion can also be applied  in this case,
with the obvious modifications: the relativistic dispersion relation
has to be replaced by the lattice one
in Eq.~\eqref{latdisprel},
and integrations over the lattice momenta runs from $-\pi/a$ to $\pi/a.$
As the calculation is analogous to the field theory calculation, we report it in Appendix \ref{latFFapp}.
Exploiting the fact that the transverse magnetization
has
 a local expression in terms of the free fermionic variables,
we can go beyond the truncated form factor expansion and obtain exact transverse DSF $S^{xx}(\omega,q)$, Eq.~\eqref{eq:sTSDSF},
at any temperature (cf. Appendix \ref{sec:xx-spin-chain} for details).

In the quantum disordered region with $\omega\ll \kB T\ll\Delta$,
the asymptotic expression for $S^{xx}(\omega)$ reads
(cf. Appendix \ref{sec:asymptotic_Sxx})
\be
S^{xx} (\omega)\approx\frac{\Delta}{\pi J^2}
e^{-\frac{\Delta}{\kB T} }
\left[ -\ln\left(\frac{\omega}{4\kB T}\right)+\frac{\kB T}{2\Delta}-\gamma_E \right]\,.
\label{asymptotic0}
\ee
Using $J=c/(2a)$ and recalling the rescaling factor $2a$ between the $\sig^x$ and the field theory operator $\eps,$ we find perfect agreement with the result Eq.~\eqref{C11asym}.

\begin{figure}[t!]
\centering
\includegraphics[width=0.47\textwidth]{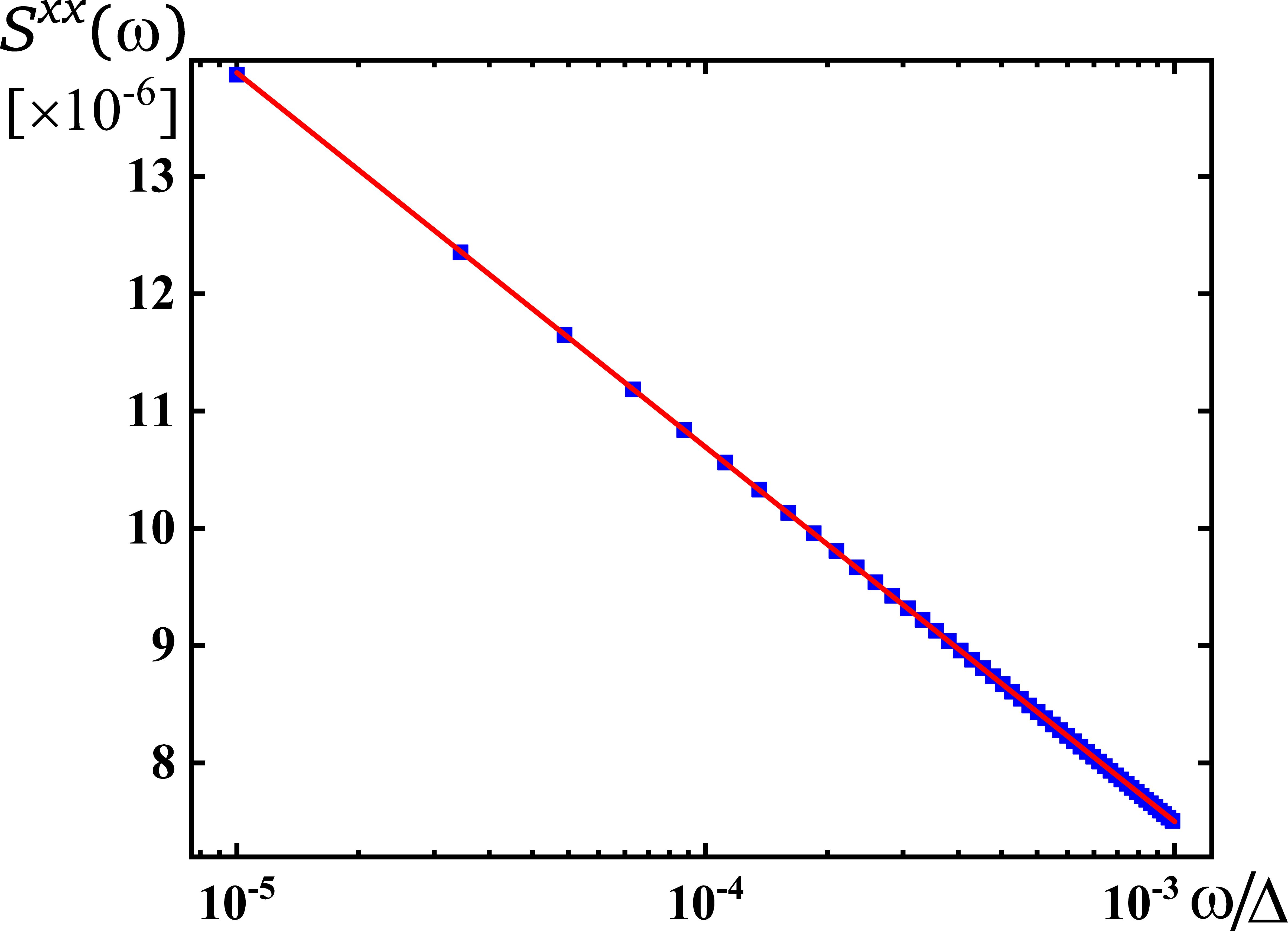}
\caption{
The local transverse DSF {\it vs.} frequency (blue dots)
at fixed $\Delta=0.1J$
and
fixed
 $\kB T=0.01J$.
The frequency dependence is fitted by
$ 10^{6}\times S^{xx}(\omega)= - 2.08 - 1.39 \ln (\omega/\Delta )  $, shown by the red solid line.
}
\label{fig:log_omega}
\end{figure}

Fig.~(\ref{fig:log_omega}) shows the frequency dependence of $S^{xx}(\omega)$,
evaluated from the lattice model, and its fit
in the $\omega\ll \kB T\ll\Delta$ region.
With parameters $\Delta=0.1J$ and $\kB  T=0.01J$,
the exact lattice result Eq.~\eqref{eq:sTSDSF}
gives the expected logarithmic divergence in $\omega$:
$
10^{6}\times S^{xx}(\omega)= - 2.08 - 1.39 \ln (\omega/\Delta ).
$
This agrees well with the asymptotic result
Eq.~\eqref{asymptotic0},
$
10^{6}\times S^{xx}(\omega)= - 1.99 - 1.38 \ln (\omega/\Delta )
$.

\begin{figure}[t!]
    \centering
   \includegraphics[width=0.48\textwidth]{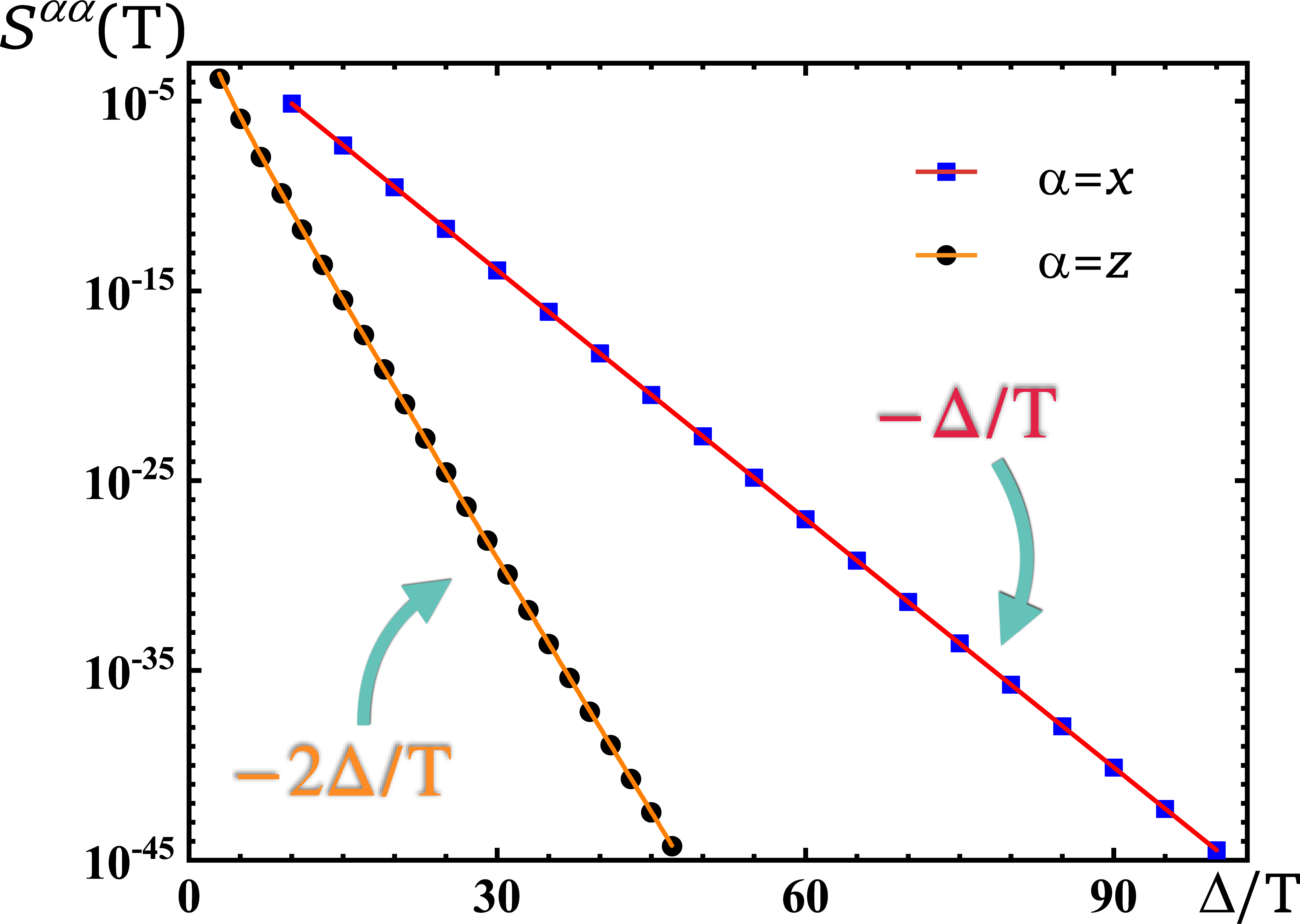}
    \caption{
The contrast of
the
thermal-activation
gaps
between $S^{xx}$ (blue dots)
and $S^{zz}$ (black dots);
 the gap for $S^{xx}$ is half of that for $S^{zz}$.
    }
    \label{Sz_beta}
\end{figure}

Similarly, we also study the temperature dependence of $S^{xx}$ with $\omega=10^{-4}J$ and $\Delta=0.1J$.
The exact lattice result,
shown in Fig.~(\ref{Sz_beta}),
exhibits an exponential decay in temperature together with a logarithmic correction
in the prefactor. A fit to the lattice result yields
$
S^{xx}(T)=e^{-\Delta/(\kB T)}\left[ 0.24 + 0.03 \ln (\kB T/\Delta)\right]
$
which conforms with the result obtained from the  asymptotic expression Eq.~\eqref{asymptotic0},
$
S^{xx}(T) =e^{-\Delta/(\kB T)}\left[0.23 +0.02 (\kB T/\Delta) + 0.03 \ln (\kB T/\Delta)\right]$.
In the fitting, the $ \kB T/\Delta$ term is not taken into account since it is
negligible compared with other terms in the considered region.


{\it Local longitudinal DSF at low frequencies.~~}
We next turn to the DSF of the order parameter
field, i.e. $S^{zz}(\omega)$.
This operator is highly nonlocal in terms of the Jordan--Wigner fermions
prohibiting an exact calculation based on free fermion techniques.
However, one can still use the truncated form factor series approach.
The calculation of the form factors of $\sigma^z$ are far from being trivial,
but are known exactly even on the finite spin chain \cite{Iorgov_2011}.
We consider the paramagnetic phase in the scaling limit,
focusing on the DSF of the continuum spin operator $\sig(x)$ in Eq.~\eqref{sbar}.
We obtain, for $\kB{T} \ll \Delta$
(cf. Appendix~\ref{sec:zzscaling} for details),
\be
S^{zz}(\w)
\approx\frac{\bar\sigma^2}\Delta\frac{3\sqrt{3}}{2\pi}\left(\frac{\kB T}\Delta\right)^2e^{-\frac{2\Delta}{\kB T}}\,,
\label{S12S21}
\ee
where $\bar \sigma = \bar s (\Delta/c^2)^{1/8}$ (cf. Eq.\eqref{sbar}).
This result agrees with the scaling limit of the corresponding result found in
Refs.~\cite{steinberg_nmr_2019,unpublished}.
In Fig.~(\ref{Sz_beta}) we show the temperature dependence of $S^{zz}(\omega)$ Eq.~\eqref{S12}
by numerical evaluation with fixed frequency $\w=0.1\Delta$ on a logarithmic scale.
The fitted line,
$-0.13571 - 1.99478 \Delta/(\kB T) - 2.06216 \ln[\Delta/(\kB T)]$,
is consistent with the prediction of Eq.~\erf{S12S21},
$-0.189959- 2 \Delta/(\kB T) - 2\ln[\Delta/(\kB T)]$.

{\it Contrasting thermal activations of $S^{xx}$ and $S^{zz}$.~~}
Our key result is the different exponential temperature dependence,
$e^{-\Delta/(\kB T)}$ vs. $e^{-2\Delta/(\kB T)}$,
for $S^{xx}$ and $S^{zz}$, respectively.
This is eventually a consequence of the $\mathbb{Z}_2:\sigma^z\to-\sigma^z$ symmetry of the model.
For large transverse fields $g\gg1,$ the ground state corresponds to spins pointing in the $x$ direction,
so intuitively, excitations in the paramagnetic phase can be thought of as spin flips in the transverse direction generated by $\sigma^z$.
This means that $\sigma^z$ creates and destroys particles that carry the same quantum numbers as the operator itself,
i.e. they are odd under spin reversal in the $z$ direction.
As a consequence, the nonzero matrix elements of $\sigma^z$ in the paramagnetic phase
are between states with different particle number parity \cite{Berg_1979,Iorgov_2011}.
For the transverse magnetization $\sigma^x$ it is the opposite:
its only nonvanishing matrix elements are between states of the same parity because it is quadratic in the fermionic operators.

As discussed in the introduction, at low frequencies $\w \ll \Delta$
only those matrix elements can contribute in the Lehmann-representation where the energies of the two states are close to each other due to the Dirac-delta expressing energy conservation.
This implies that matrix elements between the ground state and
the excited states do not contribute.
Moreover, independently of which state carries the Boltzmann factor, the contribution of the matrix element will be $\sim e^{-n\Delta/(\kB T)}$ where $n$ is the {\it larger} of the particle numbers in the two states.
Together with the parity property of $\sigma^z$ this implies that the leading temperature dependence of the longitudinal DSF is $e^{-2\Delta/(\kB T)}$ coming from 1p - 2p contributions
while that of the transverse DSF is $e^{-\Delta/(\kB T)}$ coming from 1p - 1p matrix elements.

We note in passing that in the quantum $E_8$ integrable field theory,
the scaling limit of the Ising spin chain perturbed by a {\it longitudinal} magnetic field at its QCP ($g=g_c=1$),
even the longitudinal DSF features a $e^{-\Delta/(\kB T)}$ decay \cite{jianda_E8_2014}.
This is a consequence of the fact that $\sigma^z$ has nonzero 1p - 1p matrix element \cite{Delfino_1995} as the $\mathbb{Z}_2$ symmetry is explicitly broken by the perturbing field.



{\it NMR relaxation rates.~~}
The transverse field
applied along the $x$-axis
in the TFIC
serves as the applied static magnetic field in an NMR setup \cite{kinross_evolution_2014}.
The longitudinal NMR relaxation rate
in this field geometry
is given by \cite{Moriya1956,Jaccarino1965}
\be
\frac1{T_1}
\sim
 |A_y|^2
 S^{yy}(\omega_n)+
|A_z|^2 S^{zz}(\omega_n) \, .
\ee
Here, $A_j$ ($j$ = $x$, $y$ and $z$) is the scalar hyperfine coupling,
which we take as constants for simplicity,
and $\omega_n$ is the resonant frequency ($\sim$MHz) of NMR measurements.
Therefore,
$1/T_1$
probes local spin dynamics
through $S^{zz}$ and $S^{yy}$ along the two orientations orthogonal to the transverse-field direction $x$.
In the TFIC, $S^{yy}(\omega) =S^{zz}(\omega) \omega^2/[4(g J)^2]$ \cite{jianda_E8_2014}
so the contribution from $S^{yy}$ can be ignored
in an
 NMR
setting (which is at a very low frequency).
As such, we expect
$1/T_1 \sim S^{zz} (\omega = \omega_n)
 \sim e^{-2\Delta/(\kB T)}$, for $\omega\ll \kB T\ll \Delta$ \cite{kinross_evolution_2014}.

To probe $S^{xx}(\omega)$, we consider the
measurement of  $1/T_2$
with the same field geometry as
$1/T_1$.
In general \cite{Moriya1956,Jaccarino1965}
\bea
\frac{1}{T_{2}} =&&
\frac{1}{T_{1}^{\prime}} +\frac{1}{T_2^{\prime} } \,, \nonumber\\
\frac{1}{T_1^{\prime} }  = && A \frac{1}{T_1 } \,,
\label{eq:T2rate}
\\
\frac{1}{T_2^{\prime} }  = && |A_x|^2 S^{xx} (\omega \rightarrow 0)\, . \nonumber
\eea
Here, the non-secular contribution $1/T_1^{\prime} $
can be estimated from the result of $1/T_1$
 measurement using the prefactor {\it A} calculated theoretically based on Bloch-Wangsmann-Redfield theorey, as outlined in Appendix I.
The secular term $1/T_{2}^{\prime}$
can then be determined from  Eq.\,\eqref{eq:T2rate} (first line); in practice, this separation is the easiest at relatively
low temperatures,
where the non-secular term
$1/T_1^{\prime}$
has been suppressed
by the larger gap $2\Delta$ of $S^{zz}$
but
$1/T_2^{\prime} \sim S^{xx}(\omega \rightarrow 0) 
\sim e^{-\Delta/(\kB T)}$ still remains sizable.

{\it Conclusion.~~}
To conclude, we determined
the leading behavior of the local {\it transverse}
DSF in the quantum disordered region of the TFIC at
small transfer energy
with temperature much smaller than the gap.
It is shown that when the transfer energy is
much smaller than the
temperature the local transverse DSF exhibits
a logarithmic singularity.
We found that the extracted thermal activation gap
from the local {\it transverse} DSF is half of that for the
{\it longitudinal} one, which can be attributed to the different
parities of $\sigma^x$ and $\sigma^z$ in the quantum disordered
region of the TFIC. This sharp contrast can be directly
tested in a proper NMR setup through
$1/T_1$ and $1/T_2$ relaxation rate measurements.
It is known that the thermodynamical quantity, the Gr\"uneisen
ratio of the TFIC exhibits unique quantum critical behavior \cite{jianda_crossovers_2018}.
Our work reveals a
sharp contrast of
the temperature dependence of the
{\it transverse} and {\it longitudinal} local DSFs of the TFIC,
which represents
a unique
and surprising
dynamical feature for the TFIC.
Accessing the universality of the TFIC is a crucial step
toward a realization of the exotic quantum
phenomena such as the new particles in the
quantum
$E_8$ integrable model \cite{jianda_E8_2014, zhang_e8_observation_2020, xiao_cascade_2021,zou_e_8_2021}.
Our work implies that combined measurements of
the dynamical and thermodynamic quantities
provide telltale experimental signs for
the class of TFIC universality.

\section*{Acknowledgments}
The work at Shanghai Jiao Tong University
was partially supported
by
Natural Science Foundation of Shanghai with Grant No.
20ZR1428400 and Shanghai Pujiang Program with Grant No.
20PJ1408100 (J.Y, J.W),
and by National Research, Development and
Innovation Office (NKFIH) under the research grant K-16 No. 119204 and by the Fund
TKP2020 IES (Grant No. BME-IE-NAT), under the auspices of the Ministry
for Innovation and Technology.
J.W. acknowledges additional support from a Shanghai talent
program.
The work at McMaster was supported by NSERC.
The work at Rice has been supported  in  part  by  the  NSF  Grant  No.   DMR-1920740 and  the  Robert  A.  Welch
Foundation  Grant  No.  C-1411.
Q.S. acknowledges the hospitality of the Aspen Center for Physics, which is supported by the NSF grant No.PHY-1607611.
 M.K. was supported by the \'UNKP-20-5 new National Excellence Program of the Ministry for Innovation and Technology from the source of the National Research, Development and Innovation Fund,
 and acknowledges support by a Bolyai J\'anos grant of the HAS.


\onecolumngrid

\appendix


\section{Diagonalization of the lattice and field theory Hamiltonians}
\label{sec:J-W}

After Jordan--Wigner transformation, $\sigma^x_i=1-2c^\dag_ic_i$ and
$\sigma _i^z  =  - \prod\limits_{j < i} {\left( {1 - 2c_j^\dag  c_j } \right)} \left( {c_i  + c_i^\dag  } \right)$,
the TFIC Hamiltonian Eq.~(\ref{Hamiltonian}) becomes
\be
\label{eq:H_JW_fermion}
    H_\text{I}=-J\sum_{i=1}^N\left[(c_i^\dagger c_{i+1}+c_i^\dagger c_{i+1}^\dagger+h.c.)
    +g(1-2c_i^\dagger c_i)\right]
\ee
in terms of the fermionic operators $c_i,c_i^\dag.$
%
%
%
After Fourier transformation, $c_j=\frac1{\sqrt N}\sum_k c_k e^{i k j},$
the Hamiltonian is diagonalized by a Bogoliubov rotation
$\gamma_k=u_k c_k-iv_k c_{-k}^\dagger\,,$
where
$
u_k=\cos(\theta_k/2)\,, v_k=\sin(\theta_k/2)
$
with the Bogoliubov angle $\tan(\theta_k)=\frac{\sin k}{g-\cos k}$.
 After these steps, we arrive at
\be
\label{eq:H_gamma}
H_\text{I}=\sum_k \epsilon_k\left(\gamma_k^\dagger\gamma_k-\frac{1}{2}\right)
\ee
with single-particle energy dispersion $
\epsilon_k=2J\sqrt{1+g^2-2g\cos k}.$

The field theory Hamiltonian Eq.~\eqref{HFT} can be diagonalized by the plane wave expansion of the Majorana fields which in the paramagnetic phase reads (setting $\hbar=1$)
\begin{subequations}
\label{psiexp}
\begin{align}
\psi(x,t) &= \phantom{i}\sqrt{\frac{mc}2}\int\frac{\ud\th}{2\pi} e^{-\th/2}\left[ \alpha  a(\th)e^{ip_\th x-i\eps_\th t} +  \alpha^* a^\dag(\th)e^{-ip_\th x+i\eps_\th t}\right]\,,\\
\bar\psi(x,t)&= i\sqrt{\frac{mc}2}\int\frac{\ud \th}{2\pi} \,e^{\th/2}\;\,\left[\alpha a(\th)e^{ip_\th x-i\eps_\th t} -  \alpha^*a^\dag(\th)e^{-ip_\th x+i\eps_\th t}\right]\,,
\end{align}
\end{subequations}
where $\alpha=e^{-i\pi/4},$ and $p_\th=mc\sinh(\th)$ and $\eps_\th=mc^2\cosh(\th)$ are the momentum and energy in terms of the rapidity variable $\th.$ The creation/annihilation operators obey the algebra
\be
\{a(\th),a^\dag(\th')\} =2\pi\delta(\th-\th')
\ee
and diagonalize the Hamiltonian which becomes
\be
\mathcal{H}_{\text I} = \int\frac{\ud\th}{2\pi}\,a^\dag(\th)a(\th) \,mc^2\cosh\th\,.
\ee

\section{Form-factor method and linked cluster expansion}
\label{sec:form-factor}

Exploiting the local nature of the transverse magnetization in terms of the fermions, the transverse DSF $S^{xx}(\w,q)$ can be obtained exactly. This is however not true for the longitudinal DSF $S^{zz}(\w,q).$
Still, in both cases one can give a systematic low temperature expansion \cite{Essler_2009,gabor_2010}. As we are mainly interested in the low-temperature NMR relaxation rates, we first discuss this more general approach, applicable both in the spin chain and in the field theory. We shall use the field theory notations but everything can be translated to the spin chain in a straightforward manner.

Our starting point is the Lehmann spectral representation,
%
\be
\label{lehmann}
S^{\alpha\alpha}(\w,q)=
\sum_l\int_{-\infty}^\infty\ud t\,e^{i\w t-iql}\langle \sig_{l+1}^\alpha(t) \sig^\alpha_1(0)\rangle_T=
\frac1{\mc{Z}}\sum_{n,m}e^{-\beta E_n}(2\pi)^2\delta(\hbar \w+E_n-E_m)\delta(q+P_n-P_m)
|\langle n|\sig^\alpha_1(0)|m\rangle|^2\,.
\ee
%
Using the multiparticle energy eigenstates\footnote{More generally, in interacting field theories the basis of asymptotic scattering states is used.} $|\th_1,\dots,\th_n\rangle$ in the Lehmann representation \eqref{lehmann} leads to
\be
S^{\alpha\alpha}(\w,q)=\frac1{\mc{Z}}\sum_{r,s=0}^\infty C^{\alpha\alpha}_{r,s}(\w,q)
\label{eq:series_app}
\ee
with $\alpha = x, y, z$ and
\begin{multline}
C^{\alpha\alpha}_{r,s}(\w,q)=\int\frac{\ud\th_1\dots\ud\th_r}{(2\pi)^rr!}\int\frac{\ud\th'_1\dots\ud\th'_s}{(2\pi)^ss!}e^{-\beta E_r}
\\(2\pi)^2
\delta(\hbar \w+E_r-E_s)\delta(q+P_r-P_s)
\,|\cev{\th_1\dots\th_r}\sig^\alpha(0,0)\vec{\th'_1\dots\th'_s}|^2\,,
\end{multline}
where the energy and momentum eigenvalues are
$E_n=mc^2\sum_i^n\cosh\th_i$ and $P_n=mc\sum_i^n\sinh\th_i.$

This series is a low-temperature expansion in the following sense. The term $C_{r,s}$ (we omit the $^{\alpha\alpha}$ superscript)
carries a factor $e^{-\beta E_r}<e^{-r\beta m}$, thus the small expansion parameter is $e^{-\beta m}.$ Its dependence on $s$ is
less obvious but note that thanks to the energy conserving Dirac-delta,
the energies of the two states with $r$ and $s$ particles are related. For a fixed frequency $\omega$ and at low temperature one can truncate the double sum in both $r$ and $s$. The partition function can also be written as
$\mc{Z}=\sum_{n=0}^\infty \mc{Z}_n$ where $\mc{Z}_n$ has a factor of
$e^{-\beta E_n}$.

Now the main problem to be solved is the
regularization of the singularities present in the partition function
and in the matrix elements (form factors) of the operators in question.
In infinite volume all $\mc{Z}_n$ are singular due to
the fact that they contain a scalar product of two momentum
eigenstates. Similarly, $C_{r,s}$ inherits the
kinematical poles of the form factors whenever two rapidities in the
two sets coincide, i.e. $\th_i=\th'_j$ for some $i,j$. Since the structure factor is a well-defined physical quantity, these singularities must cancel each other. In order to make this manifest we reshuffle the infinite series in a linked cluster expansion \cite{Essler_2009,gabor_2010}
\be
\chi^{\alpha\alpha}(t,x)=\sum_{r=0,s=0}^\infty D^{\alpha\alpha}_{r,s}(t,x)\,,
\ee
where the terms
\begin{align}
D_{0,s}&=C_{0,s}\,,\\
D_{1,s}&=C_{1,s}-\mc{Z}_1C_{0,s-1}\,,\\
D_{2,s}&=C_{2,s}-\mc{Z}_1C_{1,s-1}+(\mc{Z}_1^2-\mc{Z}_2)C_{0,s-2}+\dots
\end{align}
are supposed to be finite, and equivalent relations hold with the
indices interchanged. In order to obtain a finite result one needs to
regularize the divergencies either in a continuum scheme by adding
infinitesimal imaginary parts to the rapidities \cite{Essler_2009}, or by
going to a large but finite volume $L$ that satisfies $1\ll mL\ll
e^{m\beta}$, so that the density of thermally excited particles is
small \cite{Essler_2009,gabor_2010}. The singularities manifest themselves as
positive powers of $L$, while the final result for the $D_{r,s}$ should be $\sim\mc{O}(L^0)$.
All the $D_{1n}(t,x)$ terms are given for any massive relativistic
diagonal scattering theory in Ref.~\cite{gabor_2010}, while the general
expression for $D_{22}(t,x)$ can be found in Ref.~\cite{Gabor_2012}.

The resulting series still can have diverging terms, signalling that the
zero temperature quantity is already singular. This happens around the single particle
dispersion relation $\hbar \omega\sim\epsilon(k)$, where the zero temperature DSF is
proportional to an on-shell Dirac-delta which broadens at non-zero
temperatures \cite{Essler_2009}.
In these cases a resummation of infinitely many terms is necessary in order to obtain the finite result.
However, if we are interested in the small-$\w$ behavior in
the disordered phase of the Ising model then due to $\hbar \w\ll m c^2$ we are far from the mass shell.
In this case the individual terms are not singular and the truncated series should give
a good approximation.

\section{Detailed field theory calculation of local transverse DSF}

\label{app:xxscaling}

In this appendix we provide the details of the calculations leading to Eq.~\eqref{C11asym}.
Our starting expression is Eq. \eqref{C11} from the main text,
\be
    C_{11}(\w,q)=
\int\int\frac{\ud\th}{2\pi}\frac{\ud\th'}{2\pi}
|F_{2}^{\eps}(\th|\th')|^2
e^{-\beta m\cosh\th}
(2\pi)^2\delta[q+m(\sinh\th-\sinh\th')]\delta[\w+m(\cosh\th-\cosh\th')]\,.
\ee
Both integrals can be performed using the Dirac-delta constraints. The system of equations can be brought to the following form in terms of $x=e^{\th'}$ and $y=e^\th$
\begin{subequations}
\begin{align}
x-y&=\tilde\w+\tilde p\equiv A\,,\\
\frac1x-\frac1y&=\tilde\w-\tilde p\equiv B\,,
\end{align}
\end{subequations}
where $\tilde\w=\w/m,\tilde q =q/m.$ This leads to the two solutions $\{x_+,y_+\}$ and $\{x_-,y_-\}$
\begin{subequations}
\begin{align}
x_\pm &= \frac{AB\pm\sqrt{AB(AB-4)}}{2B}\,, \\
y_\pm &= \frac{-AB\pm\sqrt{AB(AB-4)}}{2B}\,.
\end{align}
\end{subequations}
These roots must be real and positive, so their product must be positive, implying
\be
x_+ y_+ = x_- y_- = -x_+ x_- = -\frac{A}B = \frac{\tilde q+\tilde w}{\tilde q-\tilde w}>0\,,
\ee
so $\tilde q^2-\tilde w^2>0.$ Their sum is also positive, leading to
\be
x_\pm+y_\pm =\frac{\pm\sqrt{AB(AB-4)}}B\,,
\ee
so only the $``+"$ solution is valid for $B=\tilde\w-\tilde q>0$ and only the $``-"$ solution is valid for $B=\tilde\w-\tilde q<0.$ Note that since $AB<0,$ the expression under the square root is always positive so the reality condition is automatically satisfied. Summarizing,
\begin{align}
e^{\th_\pm} &= \frac{q^2-\w^2\pm\sqrt{(q^2-\w^2)(q^2-\w^2+4m^2)}}{2m(\w-q)}\,,\\
e^{\th'_\pm} &= \frac{-q^2+\w^2\pm\sqrt{(q^2-\w^2)(q^2-\w^2+4m^2)}}{2m(\w-q)} \,,
\end{align}
where the $"+"$ roots are valid for $q<-|\w|$ and the $"-"$ roots are valid for $q>|\w|.$
The Jacobian of the change of variables
\be
\{m(\cosh\th'-\cosh\th), m(\sinh\th'-\sinh\th)\}\to\{\th,\th'\}
\ee
is $m^2\sinh(\th-\th').$ Conveniently,
\be
\cosh(\th-\th') = 1+(q^2-\w^2)/(2m^2)\,,\qquad  \cosh\th_{\pm} = \left(-\w\mp q\sqrt{1+4m^2/(q^2-\w^2)}\right)/(2m)\,.
\ee
Using $|F^{(\eps)}(\th|\th')|^2= m^2[1+\cosh(\th-\th')]/2$
we obtain
\be
C_{11} (\w ,q) = e^{-m\beta\cosh\th_\pm}\frac{q^2-\w^2+4m^2}{2\sqrt{(q^2-\w^2)(q^2-\w^2+4m^2)}}
= \frac12e^{\beta\w/2}\sqrt{1+\frac{4m^2}{q^2-\w^2}}e^{-\frac{\beta}2|q|\sqrt{1+\frac{4m^2}{q^2-\w^2}}}\,,
\ee
where we used $\pm q = -|q|.$
The local dynamic structure factor is given by
\be
C_{11}(\w) = \int\frac{\ud q}{2\pi} C_{11}(\w,q) = 2\int_{|\w|}^\infty\frac{\ud q}{2\pi} C_{11}(\w,q)\,.
\ee
Note that as long as $\w$ is non-zero, the exponential cuts off
the diverging prefactor and the integrand remains finite. This is no longer true for $\w=0$ which signals a logarithmic singularity.
We can extract the leading behavior in the small-$\w$ (and low-$T$) limit as
\be
\label{C11asymapp}
\begin{split}
&C_{11}(\w) \approx \frac12e^{\beta\w/2}
\int_{|\omega|}^\infty\frac{\ud q}{\pi}
\left(\frac{2 m}{q} + \frac{q}{4 m} \right)e^{-\frac{\beta m}{2}\left(\frac{q^2}{4 m^2}+2+\frac{\omega^2}{q^2}\right)}
\\&
\approx
\frac12e^{\beta\w/2}
\int_{\omega}^{p_m}\frac{\ud q}{\pi}
\left(\frac{2 m}{q} + \frac{q}{4 m} \right)e^{-\frac{\beta m}{2}\left(2 +\frac{\omega^2}{q^2}\right)}
+
\frac12e^{\beta\w/2}
\int_{p_m}^\infty\frac{\ud q}{\pi}
\left(\frac{2 m}{q} + \frac{q}{4 m} \right)e^{-\frac{\beta m}{2}\left(\frac{q^2}{4 m^2}+2\right)}
\\&
=
\frac{e^{\beta\w/2}e^{-m\beta}}{32\pi}
\left\{ \frac{2\omega}{m}  e^{-\frac{\beta  m (m+\omega )}{2 m+\omega }} \left(e^{\frac{\beta  m}{2}} (2 m+\omega )-\omega  e^{\frac{\beta  m \omega }{4 m+2 \omega }}\right)
+\left(16 m-\beta  \omega ^2\right) \left[ \Gamma \left(0,\frac{m \beta  \omega }{4 m+2 \omega }\right)-\Gamma \left(0,\frac{m \beta }{2}\right) \right]
 \right\}
\\&\approx
 -\frac{m}{\pi} e^{\beta \w/2}e^{-m\beta}\left[\ln\left(\frac{\beta\w}4\right)-\frac1{2m\beta}+\gamma_\text{E}\right]\,,
\end{split}
\ee
where
the incomplete gamma function $\Gamma (a,z) = \int_z^\infty  {{t^{a - 1}}{e^{ - t}}dt}$,
Euler's constant $\gamma_\text{E}\approx0.57712$, and $p_m= \sqrt{\omega(2 m+\omega )}$ is the extreme point of the exponent.

\subsection{Alternative derivation of the local transverse DSF}
\label{app:alt}

Another way to obtain the result in Eq. \erf{C11asym} is to focus on the {\it local} DSF from the start, defined as
\be
S^\eps(\w)=\int\ud t \,e^{i\w t}\langle \eps(0,t) \eps(0,0)\rangle_T\,.
\ee
This contains a disconnected piece proportional to $\delta(\w).$ The first nontrivial term in the expansion of the connected part is
%
\be
C_{11}(\w) = \int\frac{\ud\th}{2\pi}\int\frac{\ud\th'}{2\pi}
|F_{2}^{\eps}(\th|\th')|^2 
e^{-\beta m\cosh\th}2\pi\delta[\w+m(\cosh\th-\cosh\th')]\,.
\ee
Let us assume that $\w>0$ so the energy conservation condition has two real solutions $\th_+$ and $\th_-=-\th_+$ for all $\th,$ where we denote the positive solution by
\be
\th_+ = \mathrm{arccosh}(\cosh\th+\w/m)>0\,.
\ee
Now
%
$\delta[\w+m(\cosh\th-\cosh\th')] = \frac1{m\sinh\th_+}\left[\delta(\th'-\th_+)+\delta(\th'+\th_+)\right],$
%
so
%
\be
C_{11}(\w) =
m\int\frac{\ud\th}{2\pi}
 \frac{e^{-\beta m\cosh\th}}{\sqrt{(\cosh\th+\w/m)^2-1}}
[1+\cosh\th(\cosh\th+\w/m)]\,,
\ee
%
where we used the identity $\cosh^2[(\th-\th_+)/2]+\cosh^2[(\th+\th_+)/2] = 1+\cosh\th \cosh\th_+.$ Introducing the shorthand notation $\tilde\w=\w/m$ and changing integration variable to $u=\cosh\th,$
\be
C_{11}(\w)  =
  \frac{m}{\pi}\int_{1}^\infty\ud u\,
 e^{-m\beta u}\frac{u^2+\tilde\w u+1}{\sqrt{(u-1)(u+1)(u-1+\tilde w)(u+1+\tilde w)}}\,.
\ee
The integral is singular for $\w=0.$ To extract the small-$\w$ behavior, we can approximate the integral by
%
%
%
%
\be
\begin{split}
C_{11}(\w) & \approx \frac{m}{\pi}\int_{1}^\infty\ud u\, e^{-m\beta u}\frac{u^2+1}{(u+1)\sqrt{(u-1)(u-1+\tilde w)}}
\approx
\frac{m}{\pi}\int_0^\infty\ud v\, e^{-m\beta (1+v)}\frac{1+v/2}{\sqrt{v(v+\tilde w)}}
\\&=\frac{m}{\pi} e^{-m\beta} \left[e^{\beta\w/2}K_0\left(\frac{\beta\w}2\right)+ \frac{\sqrt\pi}{4m\beta} U(1/2,0,\beta\w)\right]\,,
\end{split}
\ee
where $K_0(x)$ is the modified Bessel function of the second kind and $U(a,b,z)$ is the confluent hypergeometric function. Expanding the result for small $\w$ we find
\be
C_{11}(\w)\approx
\frac{m}{\pi} e^{-m\beta} \left[ -\ln \left(\frac{\beta w}4\right) +\frac1{2m\beta} - \gamma_\text{E}\right]
\ee
which agrees with the result in Eq.~\eqref{C11asymapp}.

\section{ Calculations of local transverse DSF in the spin chain: Truncated form factor series method}
\label{latFFapp}

In this appendix we present the form factor calculation, analogous to that  in App. \ref{app:xxscaling}, in the spin chain.
The transverse magnetization on the lattice is given in terms of the Jordan--Wigner fermions as
\be
\sigma_j^x = 1-2c^\dag_jc_j\,.
\ee
The Hamiltonian is quadratic in these fermionic operators and it is diagonalized by going to momentum space and performing the Bogoliubov transformation
\begin{subequations}
\label{bog}
\begin{align}
\frac1{\sqrt L} \sum_{j=1}^L c_j e^{ikj}= c(k) &= \cos(\vartheta_k/2) \alpha(k) + i\sin(\vartheta_k/2) \alpha^\dag(-k)\,,\\
\frac1{\sqrt L} \sum_{j=1}^L c^\dag_j e^{ikj}= c^\dag(-k) &= i\sin(\vartheta_k/2) \alpha(k) + \cos(\vartheta_k/2) \alpha^\dag(-k)
\end{align}
\end{subequations}
with
\be
e^{i\vartheta_k}=\frac{g-e^{ik}}{\sqrt{1+g^2-2g\cos k}}=\frac{2J(g-e^{ik})}{\eps(k)}\,.
\ee
The ground state expectation value of $\sig^x_j$ is then easily calculated to be
\be
\vev{\sig^x}\equiv \langle0|\sig^x_j|0\rangle  = 1-2 \frac1L \sum_{n=1}^L \sin^2(\vartheta_k/2) = \frac1L \sum_{n=1}^L\left(1-2\sin^2(\vartheta_k/2) \right) = \frac1L \sum_{n=1}^L \cos(\vartheta_k) \longrightarrow
\int_{-\pi}^\pi \frac{\ud k}{2\pi} \cos(\vartheta_k)\,.
\ee
We can also compute the matrix elements
\begin{align}
\langle0|\sig^x_j|k,k'\rangle &\equiv \langle0|\sig^x_j \,\alpha^\dag_k\alpha^\dag_{k'}|0\rangle = \phantom{-}\frac{2i}L e^{-i(k+k')j}\sin\left(\frac{\vartheta_k-\vartheta_{k'}}2\right)\,,\\
\langle k|\sig^x_j|k'\rangle &\equiv \langle0|\alpha_k\,\sig^x_j\, \alpha^\dag_{k'}|0\rangle = -\frac2L e^{i(k-k')j}\cos\left(\frac{\vartheta_k+\vartheta_{k'}}2\right)+\delta_{k,k'}\vev{\sig^x}\,.
\end{align}


The first contributions to the local DSF in the Lehmann representation are
\begin{align}
C_{00}(\w) &= 2\pi\delta(\w)\vev{\sig^x}^2\,,\\
C_{11}(\w) &= \sum_{n,m}e^{-\beta\eps_n} 2\pi\delta(\w+\eps_n-\eps_m) \left[ \delta(\w)\vev{\sig^x}^2 \delta_{n,m}+\left(\frac2L\right)^2\cos^2\left(\frac{\vartheta_{k_n}-\vartheta_{k_m}}2\right)-\delta_{n,m}\frac4L\cos\left(\frac{\vartheta_{k_n}-\vartheta_{k_m}}2\right)\right]\,,
\end{align}
where $\eps_n=\eps(k_n)$ with $k_n=2\pi n/L,$ $n=-L/2+1,\dots L/2$ (Ramond sector).
When calculating $D_{11}(\w),$ the first term in $C_{11}(\w)$ cancels exactly $\mc{Z}_1C_{00}(\w)$. The remaining terms containing $\delta_{n,m}$ contribute to the $\w=0$ response. Let us focus on the first nontrivial term contributing at finite $\w,$
\begin{multline}
F(\w)\equiv  \left(\frac2L\right)^2\sum_{n,m}e^{-\beta\eps_n} 2\pi\delta(\w+\eps_n-\eps_m) \cos^2\left(\frac{\vartheta_{k_n}-\vartheta_{k_m}}2\right)\\
\longrightarrow \;\;4\int_{-\pi}^\pi \frac{\ud k}{2\pi}\int_{-\pi}^\pi \frac{\ud k'}{2\pi} e^{-\beta\eps(k)} 2\pi\delta(\w+\eps(k)-\eps(k')) \cos^2\left(\frac{\vartheta(k)-\vartheta(k')}2\right)\,.
\end{multline}
The energy Dirac-delta can only be satisfied if
\be
\eps_\text{min} = 2J|1-g|\le \eps(k)+\w \le 2J(1+g) = \eps_\text{max}\,,
\ee
which implies
\be
\cos k_0\equiv -1+\frac{1+g}g\tilde\w-\frac{\tilde\w^2}{2g} \le \cos k \le 1+\frac{|1-g|}g\tilde\w-\frac{\tilde\w^2}{2g}\,,
\ee
where $\tilde\w\equiv \w/(2J).$ Note that for $\w<0$ the first inequality is automatic, while for $0<\w<2\Delta$ the second inequality is automatically satisfied. We focus on the latter case from now on. Then $\cos k'$ is given by
\be
\cos k' = \frac{1+g^2-(\w+\eps(k))^2/(4J^2)}{2g}
= \cos k -\frac{\tilde\w}g \sqrt{1+g^2-2g\cos k}-\frac{\tilde\w^2}{2g}
\ee
which has two solutions which we denote by $k'_1>0$ and $-k'_1.$ Changing variables and performing the integration over $k'$ we obtain
\be
F(\w) = \frac2{\pi} \int_{-k_0}^{k_0} \ud k\, e^{-\beta\eps(k)} \frac{\eps(k'_1(k))}{4J^2g\sin(k'_1(k))}\left[\cos^2\left(\frac{\vartheta(k)-\vartheta(k'_1(k))}2\right) + \cos^2\left(\frac{\vartheta(k)+\vartheta(k'_1(k))}2\right)\right]\,.
\ee
Now we use the identity $\cos^2[(x-y)/2]+\cos^2[(x+y)/2] = 1+\cos(x) \cos(y),$ the explicit forms of $\eps(k)$ and
\be
\cos \vartheta(k) = \frac{g-\cos k}{\sqrt{1+g^2-2g\cos k}}
\ee
together with $\eps(k'_1)=\eps(k)+\w$ to arrive at
\be
F(\w) = \frac1{Jh\pi} \int_{-k_0}^{k_0} \ud k\,e^{-\beta\eps(k)} \frac{2\tilde\eps(k)(\tilde\eps(k)+\tilde\w)-\sin^2k-\tilde\w/g\cos k(\tilde\eps(k)+\tilde\w/2)+\tilde\w^2/2}{\tilde\eps(k)\sqrt{1-(\cos k-\tilde \w/g\tilde\eps(k)-\tilde\w^2/(2g))^2}}\,,
\ee
where $\tilde\eps(k)=\eps(k)/(2J).$
We would like to obtain an approximate analytical result in the $\w\to0$ limit. Then we can keep only the $O(\tilde\w^0)$ terms in the numerator. Changing variables to $u=\cos k,$
\be
\begin{split}
F(\w) &\approx \frac2{Jg\pi} \int_{-1+(1+g)/g\,\tilde\w-\tilde\w^2/(2g)}^1 \ud u \,e^{-\beta 2Je(u)}\frac{2e(u)^2-(1-u^2)}{\sqrt{1-u^2}e(u)\sqrt{1-(u-\tilde\w e(u)/g-\tilde w^2/(2g))^2}}
\\&
\sim\frac4{\pi}\frac{|1-g|}{2Jg}e^{-\beta\Delta} \ln\tilde\w
=\frac1\pi\frac{\Delta}{J^2g}e^{-\beta\Delta} \ln\tilde\w
\end{split}
\ee
with $e(u)=\sqrt{1+g^2-2gu}.$
It's not surprising that the same logarithmic divergent behavior appears as in Eq.~\eqref{asymptotic} and Eq.\erf{C11asym}.
Comparing to the field theory we have to keep in mind that between $\eps$ and $\sig^x$ there is a rescaling factor of $-2a=-\hbar c/J$ which leads to a perfect match of the prefactors of $\ln\omega.$

\section{Transverse DSF \texorpdfstring{$S^{xx}(q,\w)$}{} in the spin chain}
\label{sec:xx-spin-chain}

In this section we discuss the transverse DSF in the spin chain.
In this section, we report the calculation of the exact transverse DSF and specify its low temperature and low frequency behavior at the end of the calculation.
The analogous derivation in the scaling limit can be found in Appendix \ref{app:FTexact}.

The transverse DSF follows by

\be
\begin{split}
    S^{xx}(\omega,q)&=\sum_{l=1}^N
    \int^\infty_{-\infty} \ud t e^ {i\omega t-i q l a}
    \cdot\left[ \left\langle\sigma^x_l(t)\sigma^x_{0}\right\rangle_T - \langle \sigma^x_l(t) \rangle_T \langle \sigma^x_{0} \rangle_T \right]\,,
\end{split}
\ee
where $\hbar\omega$ and $\hbar q$ are transferred energy and momentum, respectively.
In the following we set the lattice spacing as $a=1$ and $\hbar = 1$.
Starting from $\sigma^x_i=1-2c^\dag_ic_i$, going to momentum space and performing the Bogoliubov rotation, the thermal expectation values can be calculated in a straightforward way using the Hamiltonian Eq.~\eqref{eq:H_gamma} and Wick's theorem. The resulting expression for $S^{xx}(\omega,q)$ reads
\be
\begin{split}
S^{xx}(\omega,q)
&=\frac{4}{N}\sum_k
\left[
2\pi\delta(\omega-\epsilon_{k+q/2}+\epsilon_{k-q/2})\cdot
(u_{k+q/2}^2u_{k-q/2}^2-u_{k+q/2}v_{k+q/2}u_{k-q/2}v_{k-q/2})\cdot
(1-n_{k+q/2})n_{k-q/2}
\right.\\&\left.\quad
+
2\pi\delta(\omega+\epsilon_{k+q/2}+\epsilon_{k-q/2})\cdot
(u_{k-q/2}^2v_{k+q/2}^2+u_{k+q/2}v_{k+q/2}u_{k-q/2}v_{k-q/2})\cdot
n_{k+q/2}n_{k-q/2}
\right.\\&\left.\quad
+
2\pi\delta(\omega-\epsilon_{k+q/2}-\epsilon_{k-q/2})\cdot
(u_{k+q/2}^2v_{k-q/2}^2+u_{k+q/2}v_{k+q/2}u_{k-q/2}v_{k-q/2})\cdot
(1-n_{k+q/2})(1-n_{k-q/2})
\right.\\&\left.\quad
+
2\pi\delta(\omega+\epsilon_{k+q/2}-\epsilon_{k-q/2})\cdot
(v_{k+q/2}^2v_{k-q/2}^2-u_{k+q/2}v_{k+q/2}u_{k-q/2}v_{k-q/2}) \cdot
n_{k+q/2}(1-n_{k-q/2})
\right]
\\&=
\int^\pi_{-\pi}\ud k
\left[
2({1+f(k,q)})
\cdot\delta(\omega-\epsilon_{k-q/2}+\epsilon_{k+q/2})\cdot (1-n_{k-q/2})n_{k+q/2}
+({1-f(k,q)})
\right.\\&\left.
\cdot (
\delta(\omega-\epsilon_{k-q/2}-\epsilon_{k+q/2})
\cdot (1-n_{k-q/2})(1-n_{k+q/2})
+\delta(\omega+\epsilon_{k-q/2}+\epsilon_{k+q/2})\cdot n_{k-q/2}n_{k+q/2}
)\right]
\,
\end{split}\label{TSDSF}
\ee
with Fermi distribution function
$n_{k}=\left[ e^{ \epsilon_k/(\kB T) } + 1 \right]^{-1}$ and
\be
\begin{split}
f(k,q)&=
4J^2
\left( {[g-\cos(k-q/2)][g-\cos(k+q/2)]}
-{\sin(k-q/2)\sin(k+q/2)}  \right)
/(\epsilon_{k-q/2}\epsilon_{k+q/2})\,.
\end{split}
\ee
Here we have taken advantage of particle and energy conservation during the derivation,
for example
$
\int^\infty_{-\infty}\ud t e^{i \omega t}
\langle e^{iHt} \gamma_{k-q/2}\gamma_{k+q/2}^\dagger e^{-iHt}\gamma_{k+q/2}\gamma_{k-q/2}^\dagger \rangle_T
=2\pi\delta(\omega+\epsilon_{k+q/2}-\epsilon_{k-q/2})(1-n_{{k-q/2}})n_{{k+q/2}}
.
$
Focusing on $\omega > 0$ region,
after integration, we obtain
\be
\begin{split}
S^{xx} (\omega,q)&=\sum_{s=\pm}
\left[
2({1+f(k_s,q)})\cdot \frac{(1-n_{k_s-q/2})n_{k_s+q/2}}{|D^{-+}_{k_{s}}(\omega,q)|}
+({1-f(k_s,q)})\cdot \frac{(1-n_{k_s-q/2})(1-n_{k_s+q/2})}{|D^{--}_{k_{s}}(\omega,q)|}
\right]\,,
\end{split}\label{eq:sTSDSF}
\ee
where $k_{\pm}$ are solutions of energy conservation constraint
\be
\cos(k_\pm)=\frac{\omega ^2 \cos \left(\frac{q}{2}\right)}{16 J^2 g \sin ^2\left(\frac{q}{2}\right)}\pm \left(\frac{\omega ^2}{16 J^2  \sin ^2\left(\frac{q}{2}\right)}-1\right)^{1/2} \left(\frac{\omega ^2}{16 J^2  g^2 \sin ^2\left(\frac{q}{2}\right)}-1\right)^{1/2}\,,
\ee
and the Jacobians are
\be
|D^{--}_{k_{s}}(\omega,q)|
=\left| 4J^2g\left( \frac{\sin \left(k_s+\frac{q}{2}\right)}{\epsilon_{k_s+q/2}}+\frac{\sin \left(k_s-\frac{q}{2}\right)}{\epsilon_{k_s-q/2}} \right) \right|,\qquad s=\pm
\ee
\be
|D^{-+}_{k_{s}}(\omega,q)|
=\left| 4J^2g\left( \frac{\sin \left(k_s+\frac{q}{2}\right)}{\epsilon_{k_s+q/2}}-\frac{\sin \left(k_s-\frac{q}{2}\right)}{\epsilon_{k_s-q/2}} \right) \right|,\qquad s=\pm\,.
\ee
\begin{figure}[t]
    \centering
    \includegraphics[width=0.9\textwidth]{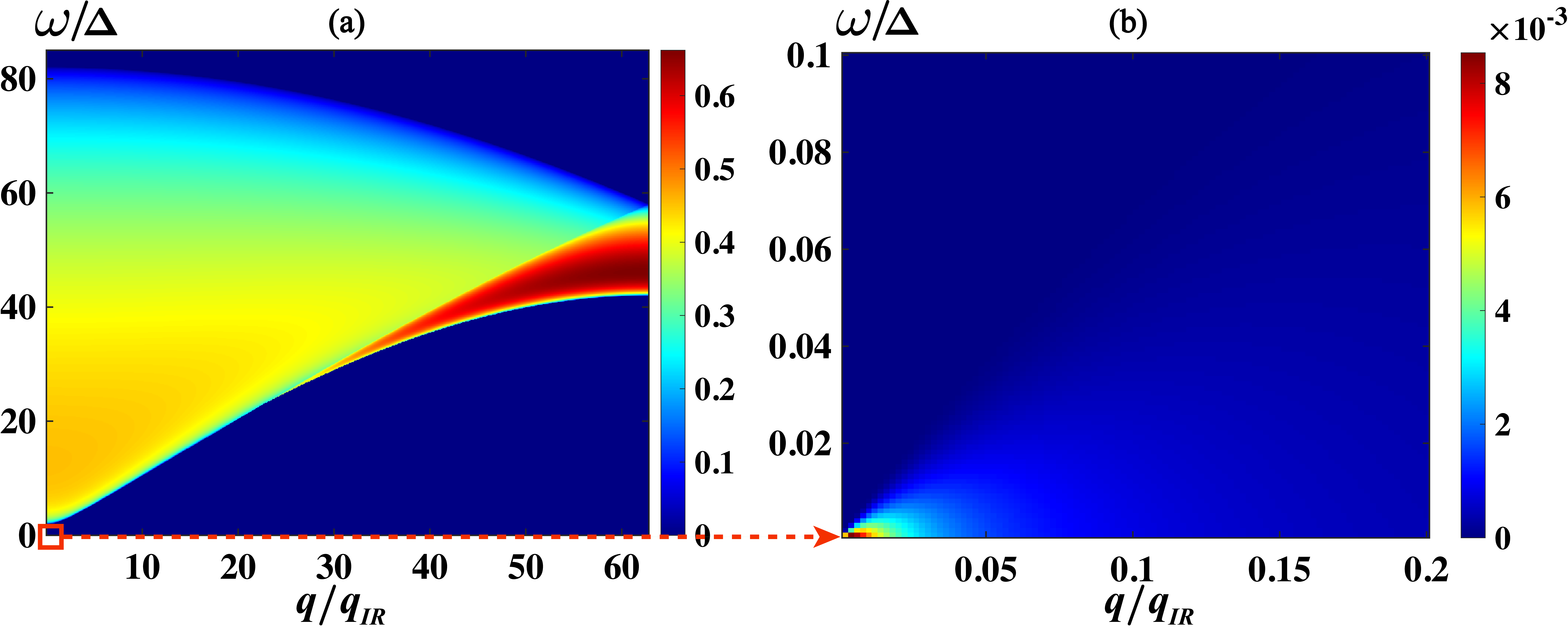}
    \caption{
    (a) The transverse DSF ${(J/\hbar)}S^{xx}(\omega, q)$ for $\omega > 0$ at $\Delta=0.1J$, $\kB T=0.01J$.
        {$q_{{}_\text{IR}}/a=\Delta/(\hbar c)=(g-1)/a$ is the infrared wave number scale.}
    The continuum above $2\Delta$ is contributed from the second term of Eq.~\eqref{eq:sTSDSF}.
    (b) The enlarged view of the small rectangular region in (a) exhibits  $S^{xx}(\omega, q)$ in the small momentum and low energy region which
    comes from the first term in Eq.~\eqref{eq:sTSDSF}.
     }
   \label{fig:DSF}
\end{figure}
The transverse DSF, measured in units of $\hbar/J$, is plotted in Fig.~\ref{fig:DSF} as a function of dimensionless variables using the infrared frequency and wave number scales $\Delta/\hbar$ and $\Delta a/(\hbar c)$, respectively.
The upper threshold in Fig.~\ref{fig:DSF} (a) is given by 
\be
\omega_\text{up}(q)=
2\epsilon_{q/2\pm\pi}=4 J \sqrt{g^2+2 g \cos \left(q/2\right)+1}\,,
\ee
while the lower thresholds are  
\begin{align}
\omega_\text{low1}(q)&=
2\epsilon_{q/2}=4 J \sqrt{g^2-2 g \cos \left(q/2\right)+1}\,,\\
\omega_\text{low2}(q)&=
\epsilon_0+\epsilon_{q}=2 J \left(\sqrt{g^2-2 g \cos q +1}+ |g-1|\right)\,.
\end{align}
In Fig. \ref{fig:DSF}~(b), the thresholds are
$\w_\text{low'}(q) = \epsilon_{q/2+k}$ and
$\w_\text{up'}(q) = \epsilon_{q/2-k}$ with $k=\arccos\left[{\cos \left(q/2\right)}/{g}\right].$

\subsection{Low temperature behavior of the local transverse DSF}
\label{sec:asymptotic_Sxx}

We now focus on the local transverse DSF in the quantum disordered region
with gap much larger than the temperature, i.e. $\kB T\ll\Delta$.
The leading contribution in Eq.~\eqref{eq:sTSDSF} is of the order
$e^{-\Delta/(\kB T)}$, and is given by the first term of Eq.~\eqref{eq:sTSDSF}.
Then the local transverse DSF follows immediately,
\bea
S^{xx} (\omega) &= &\int^\pi_{-\pi}\frac{d q}{2\pi}S^{xx}(\omega,q)=
\int^\pi_{q_c}\frac{d q}{\pi}S^{xx}(\omega,q)
\approx
\int_{q_c}^\pi \frac{d q}{\pi} \sum_{s=\pm}
2[(1+f(k_s,q)]
\frac{(1-n_{k_s-q/2})n_{k_s+q/2}}{|D^{-+}_{k_{s}}(\omega,q)|}
\\&\approx&
\int_{q_c}^{\pi}\frac{d q}{\pi}\
\left| \frac{\sqrt{1 + g^2-2 g \cos \frac{q}{2}}}{ g J \sin \frac{q}{2}} \right|
\,e^{-\epsilon_{k_-+q/2}/(\kB T)}\,, \label{sxx}
\eea
where $q_c\approx{\omega}/{2J}$ is the lower bound obtained from $\w=\w_\text{up}(q)$ at $\w \to 0$ limit.
The asymptotic behavior of the integral in the $\omega\ll \kB T\ll\Delta$ regime is determined in Appendix \ref{app:asympt} with the result
\be
S^{xx} (\omega)\approx
\frac{ e^{
-\frac{\Delta}{\kB T}}}{\pi}
\left[
-\frac{2 \Delta}{J (\Delta +2 J)}\left(\ln \left(\frac{\omega}{4\kB T} \right)+\gamma_E \right)
+
\frac{\Delta ^2+12 J^2+6 \Delta  J}{6   J^2 (\Delta +2 J)^2}\kB T
\right]\,.
\label{asymptotic}
\ee
The asymptotic result Eq.~\eqref{asymptotic} shows that finite temperature local transverse DSF
diverges logarithmically as $\omega\to0$. The energy conservation constraints in Eq.~\eqref{eq:sTSDSF}
implies only the first term of Eq.~\eqref{eq:sTSDSF} can contribute to such low-energy behavior.
Furthermore, the energy conservation leads to a constraint for the phase space. After integration of
Eq.~\eqref{TSDSF}, the constraint gives rise to the $1/q$ dependency in the integrand of Eq.~(\ref{sxx})
for small $q$, which is just the case for the lower bound dependent on the frequency.
This finally results in the logarithmic behavior.
The temperature dependence shows an exponential decay together with a logarithmic correction in the prefactor.
In the scaling limit we obtain
\be
S^{xx} (\omega)\approx\frac{\Delta}{\pi J^2}
e^{-\frac{\Delta}{\kB T} }
\left[ -\ln\left(\frac{\omega}{4\kB T}\right)+\frac{\kB T}{2\Delta}-\gamma_E \right]\,.
\ee
Using $J= c/(2a)$ and recalling the rescaling factor $2a$ between the $\sig^x$ and the field theory operator $\eps,$ we find perfect agreement with the result \erf{C11asym}.

In the $\kB T\ll\omega\ll\Delta$ region, we can simply approximate the integral by the steepest descent method and obtain the asymptotic result
\be
S^{xx}(\omega)\approx\frac{1}{\pi}
e^{-\Delta/(\kB T)}\sqrt{\frac{\pi \kB T}{\omega}}
 \frac{2\Delta}{J(2J+\Delta)}\,.
 \ee

\begin{figure}[h]
    \centering
    \includegraphics[width=0.4\textwidth]{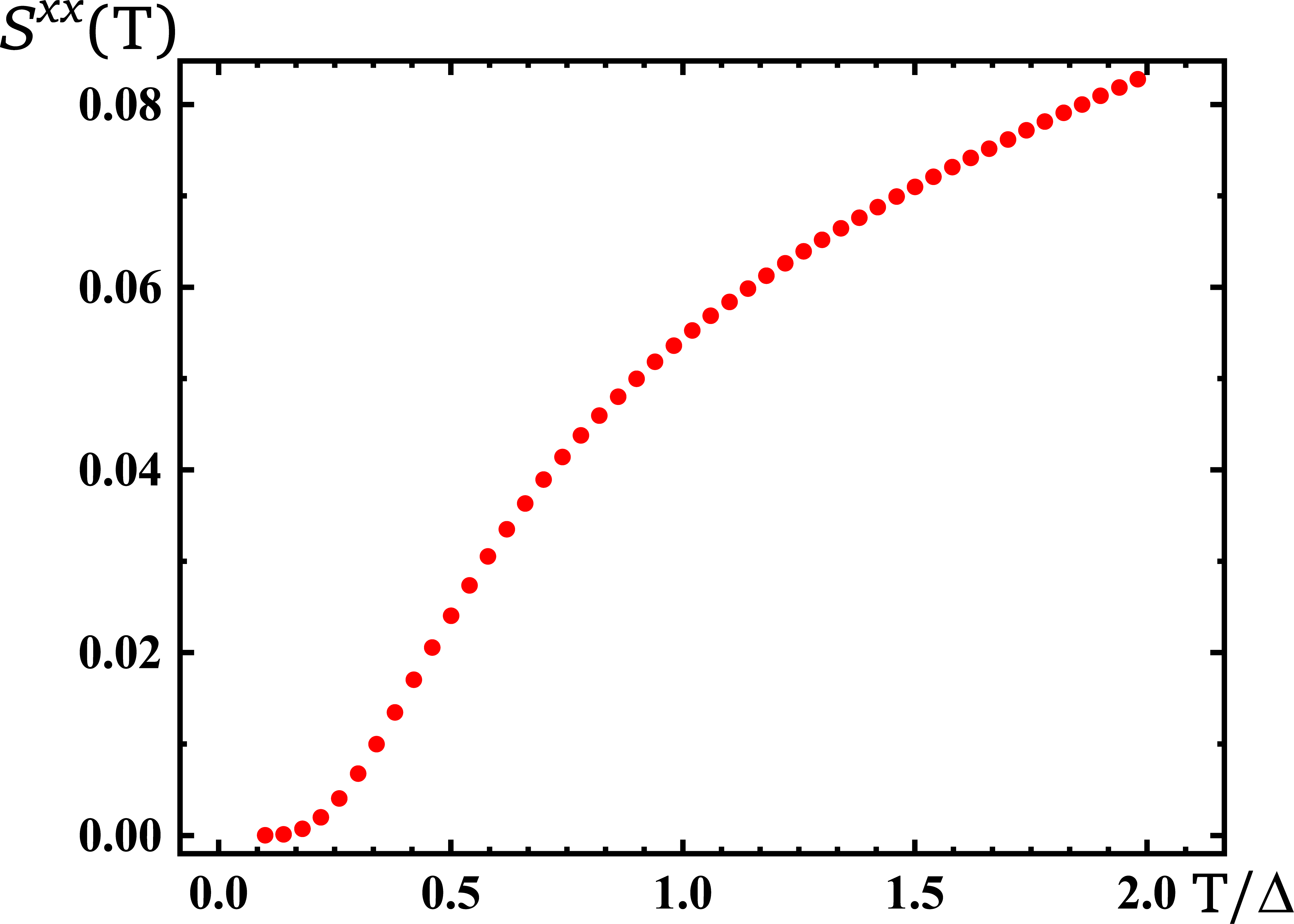}
    \caption{
    The local transverse DSF as a function of temperature at fixed $\Delta=0.1J$, and $\omega=10^{-4}J$.
    In the high $T$ region it clearly deviates from the exponentially decaying behavior.
    }
    \label{llog_beta}
\end{figure}


\section{Detailed field theory calculation of longitudinal DSF}
\label{sec:zzscaling}

In this section we turn to the DSF of the order parameter
field.
This operator is highly nonlocal in terms of the Jordan--Wigner fermions prohibiting an exact calculation based on free fermion techniques. However, one can still use the truncated form factor series approach.
This approach has been used in the study of local spin DSF and NMR relaxation rate $1/T_1$ in Refs.\,\cite{jianda_E8_2014,steinberg_nmr_2019}.
The calculation of the form factors of $\sigma^z$ are far from being trivial, but are known exactly even on the finite spin chain \cite{Iorgov_2011}.
Here we perform
the
calculation in the paramagnetic phase
in the scaling limit, focusing on the DSF of the continuum spin operator $\sig(x)$ in Eq. \eqref{sbar}.
In the disordered phase, the $\sig(x)$ operator creates and destroys particles, so its only non-zero matrix elements are between states with particle numbers of different parity, that is, the total number of particles in the two states must be odd.
The vacuum form factors are given by \cite{Berg_1979}
\be
F^{\sigma}_n(\th_1,\dots,\th_n)=\langle0|\sigma|\th_1,\dots,\th_n\rangle=
\bar\sigma\sum_{i<j}\tanh\left(\frac{\th_i-\th_j}2\right)
\ee
with $\bar\sigma=\bar s m^{1/8}$ where $\bar s$ is defined in Eq. \eqref{sbar}
and we work with $\hbar=c=1$ for the following field theory calculation.
All other matrix elements can be obtained by the crossing relation.
For example,
\be
\langle\th|\sigma|\th_1,\dots,\th_n\rangle = \langle0|\sigma|\th+i\pi,\th_2,\dots,\th_n\rangle\,,
\ee
whenever $\th\neq\th_j$ which will be the case in
our calculations.

Thus the first contributions to the DSF come from $D_{01}(t,x)=C_{01}(t,x)$ and $D_{10}(t,x)=C_{10}(t,x)$, yielding
%
\be
S^{zz}_{01}(\w,q)=\bar\sigma^2\int\frac{\ud\th}{2\pi}(2\pi)^2\delta(\w-m\cosh\th)\delta(q-m\sinh\th) \,,
\ee
and $S_{10}(\w,q)=e^{\beta\w} S_{01}(-\w,-q).$
Due to the energy conserving Dirac-delta, both $S_{10}$ and $S_{01}$ are zero for $\w<m$. It is clear that all
$D_{0,s}$ and $D_{r,0}$ will also give zero contribution, which reflects the fact that
the zero temperature result is identically zero.

Energy conservation at small frequencies also leads to a great simplification in higher orders, similarly to the case of the transverse magnetization in the previous sections.
Because of the Dirac-delta and $\w\approx0$ the two states in each
matrix element must have almost equal energies, so
$S_{r,s}(\w\approx0,q)\sim e^{-\mathrm{max}(r,s)\cdot\beta m}$. This
implies that the classification in terms of orders of $e^{-\beta m}$ is simplified, because in every order there is only a finite number of terms. For instance, in the second order one has $S_{12}+S_{21}$, in the third order $S_{23}+S_{32}$, in the fourth $S_{14}+S_{41}+S_{34}+S_{43}$, and so on.

Thus up to the second order one needs only two terms, $S_{12}$ and $S_{21}$. We use the expression for $D_{12}$ given in Ref.~\cite{Essler_2009} that can be shown to be equivalent to the more general formula in Ref.~\cite{gabor_2010},
\be
\label{D12Ess}
\begin{split}
    D_{12}&=
\frac{1}{2}\int\frac{\ud\th}{2\pi}
\int_{C_+}\int_{C_-}\frac{\ud\th_1}{2\pi}\frac{\ud\th_2}{2\pi}
|F_{3}^{\sigma}(\th+i\pi,\th_1,\th_2)|^2
e^{-\beta m\cosh\th}
\\&\times
2\pi\delta[q+m(\sinh\th-\sinh\th_{1}-\sinh\th_{2})]
2\pi\delta[\w+m(\cosh\th-\cosh\th_{1}-\cosh\th_{2})]\\
&-\bar\sigma^2\int\frac{\ud\th}{2\pi}e^{-m\beta\cosh\th}2\pi\delta(\w-m\cosh\th)2\pi\delta(q-m\sinh\th)\,,
\end{split}
\end{equation}
where the contours $C_{\pm}$ are running above and below the real axis, respectively, to avoid the kinematical poles of the form factors. But $\th=\th_i$ $(i=1,2)$ is impossible for $\w<m$, so the integrals avoid the poles even for real rapidities and there is no need to shift the contours off the real axis.
The last term is proportional to $D_{01}$ so it does not contribute for $\w<m$ and
we are left with
\be
\begin{split}
S^{zz}_{12}(\w,q)&=
\frac{1}{2}\int\frac{\ud\th}{2\pi}
\int\frac{\ud\th_1}{2\pi}\int\frac{\ud\th_2}{2\pi}
|F_{3}^{\sigma^z}(\th+i\pi,\th_1,\th_2)|^2
e^{-\beta m\cosh\th}\\
&\times2\pi\delta[q-m(\sinh\th_{1}+\sinh\th_{2}-\sinh\th)]
2\pi\delta[\w-m(\cosh\th_{1}+\cosh\th_{2}-\cosh\th)]\,.
\end{split}
\ee

\begin{figure}
\centering
\includegraphics[width=0.45\textwidth]{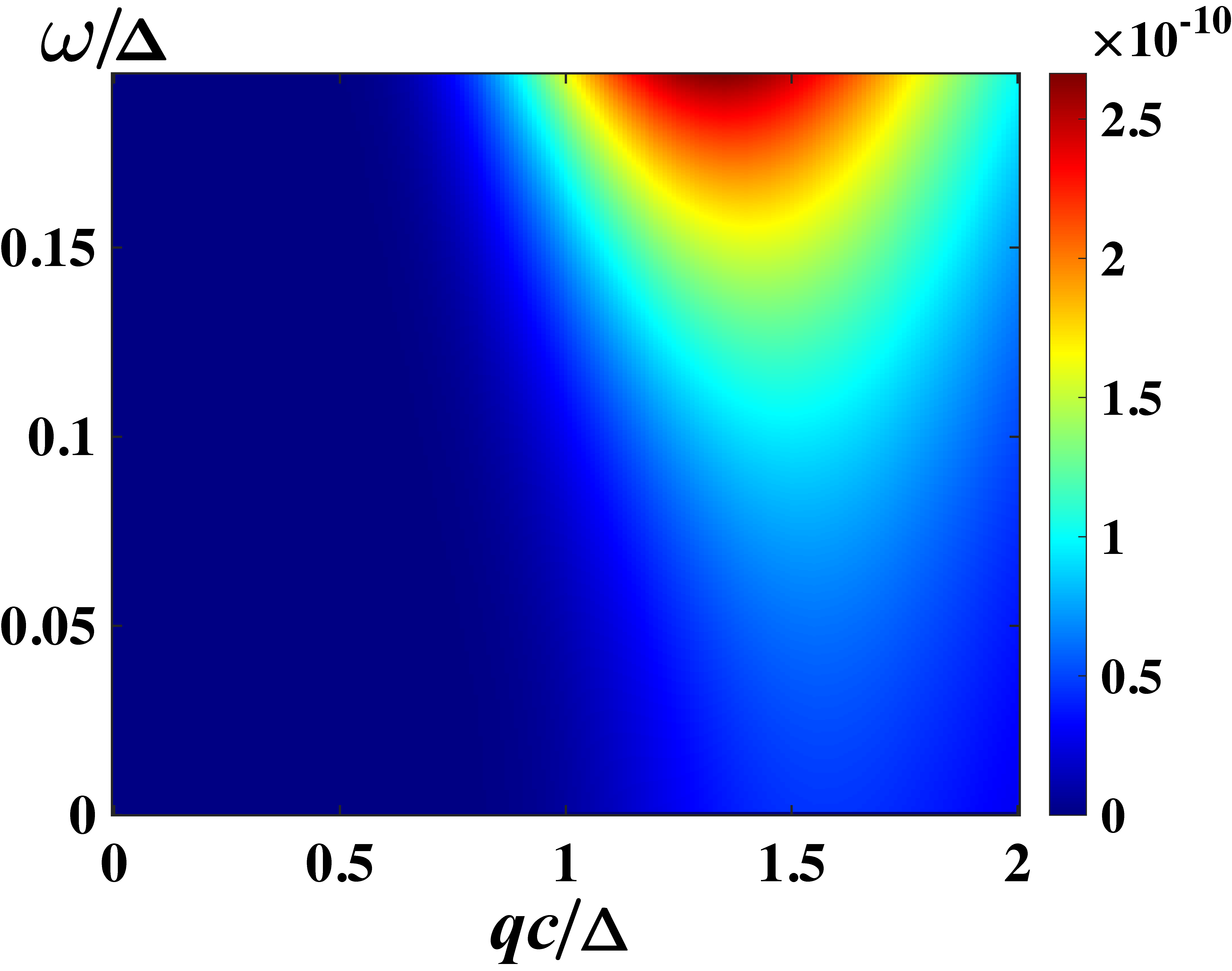}
\caption{
The leading $\mathcal{O}[e^{-2\Delta/(\kB T)}]$ contribution to the longitudinal dynamic structure factor in the scaling limit, $\Delta^2/(\hbar^2 c \,\bar\sig^2)S^{zz},$ at $\kB T=\Delta/10.$
}
\label{Szz}
\end{figure}

Exploiting the Dirac-deltas we perform the integrals over rapidities $\th_{1,2}.$ The Jacobian of the transformation is $m^2\sinh(\th_1-\th_2).$ The set of two constraint equations coming from the Dirac-deltas  has two solutions, $\{\th_1,\th_2\}=\{\th_+,\th_-\}$ and $\{\th_1,\th_2\}=\{\th_-,\th_+\}$, where
\be
\th_\pm=\log\left[\frac12\left(\w_\th+q_\th \pm \frac{\sqrt{(\w_\th^2-p_\th^2)(\w_\th^2-q_\th^2-4)}}{\w_\th-q_\th}\right)\right]
\ee
with $q_\th\equiv q/m+\sinh\th, \w_\th\equiv\w/m+\cosh\th.$ The rapidities $\th_{1,2}$ must be real which gives restrictions on the remaining rapidity $\th.$ The reality condition of $\th_{1,2}$ is equivalent to the condition that $e^{\th_1}+e^{\th_2}$ and $e^{\th_1}\,e^{\th_2}$ must be positive, which gives $\w_\th+q_\th>0,\,\w_\th-q_\th>0.$
Moreover, the combination under the square root must also be positive, $\w_\th^2-q_\th^2>4.$ One of the first two conditions, e.g. $\w_\th-q_\th>0$ can then be dropped which leaves us with two conditions. The solution of $\w_\th^2-q_\th^2>4$ for $|\w|<m$ is the following:
\begin{subequations}
\label{thdomain}
\begin{alignat}{3}
\th&<\th^{(-)}&&&  \qquad |\w|&<q\,,\\
\th&>\th^{(+)} &&& \qquad -|\w|&>q\,,\\
\th&<\th^{(-)}\;\text{ or }\;\th>\th^{(+)} &&& \qquad -\w&<q<\w\,,
\end{alignat}
\end{subequations}
and for $\w<q<-\w$ ($\w<0$) there is no solution. Here
\be
\th^{(\pm)} =  \log\left[\frac{q^2-\w^2+3m^2 \pm \sqrt{(q^2-\w^2+m^2)(q^2-\w^2+9m^2)}}{2m(\w-q)}\right].
\ee
It turns out that the other condition, $\w_\th+q_\th>0$, is automatically satisfied, so Eqs.~\erf{thdomain} give the integration domain of $\th$ in the various cases depending on $\w$ and $q.$
Thus we find
\be
S^{zz}_{12}(\w,q)=
2\int_D\frac{\ud\th}{2\pi}
\frac{e^{-\beta m\cosh\th}|F_{3}^{\sigma}[\th+i\pi,\th_+(\th),\th_-(\th)]|^2}{m^2\sqrt{(\w_\th^2-q_\th^2)(\w_\th^2-q_\th^2-4)}}\,,
\ee
where $D$ denotes the domain given in Eqs.~\erf{thdomain} and we used that $\sinh(\th_1-\th_2) = \sqrt{(\w_\th^2-q_\th^2)(\w_\th^2-q_\th^2-4)}/2.$
It is easy to see that $S_{21}(\w,q) = e^{\beta\w}S_{12}(-\w,q),$ so we have the total leading $\mathcal{O}(e^{-2m\beta})$ contribution to the DSF.
The result $S^{zz}_{12}(\w,q)+S^{zz}_{21}(\w,q)$ is plotted in Fig. \ref{Szz}.

The corresponding local DSF reads
\be
\begin{split}
S^{zz}_{12}(\w)&=
\frac{1}{2}\!\int\!\frac{\ud\th}{2\pi}
\!\int\!\frac{\ud\th_1}{2\pi}\!\int\!\frac{\ud\th_2}{2\pi}
e^{-\beta m\cosh\th}
|F_{3}^{z}(\th+i\pi,\th_1,\th_2)|^2
2\pi\delta[\w-m(\cosh\th_{1}+\cosh\th_{2}-\cosh\th)]\\&
=\frac12\int\frac{\ud\th_1}{2\pi}\int\frac{\ud\th_2}{2\pi}
\frac{e^{\beta\w-\beta m(\cosh\th_1+\cosh\th_2)}}{\sqrt{(m\cosh\th_1+m\cosh\th_2-\w)^2-m^2}}
\left[|F_{3}^{\sigma}(\th_0+i\pi,\th_1,\th_2)|^2+\{\th_0\to-\th_0\}\right]\,,
\label{S12}
\end{split}
\ee
where $\th_0=\mathrm{arccosh}(\cosh\th_1+\cosh\th_2-\w/m)>0.$
At the second order we also need $S_{21}(\w) = e^{\beta\w}S_{12}(-\w).$

We can give approximate expressions for $S_{12}$ and $S_{21}$. For
$\beta m\gg1$ only a small region around the origin in the
$(\th_1,\th_2)$ plane contributes, so we can expand both the exponent and the rest of the integrand to
second order using the explicit form factors. Performing the resulting Gaussian integrals, and expanding the result  in $\w/m$ (with $T/m\ll1$) we obtain the result in Eq. \eqref{S12S21}:
\be
\begin{split}
S^{zz}(\w)\approx
S^{zz}_{12}(\w)+S^{zz}_{21}(\w)
&\approx\frac{\bar\sigma^2}m \frac{3\sqrt{3}}{4\pi}\left(\frac1{m\beta}\right)^2e^{-2m\beta}\left[e^{\beta\w}\left(1+2\frac\w{m}\right)+\left(1-2\frac\w{m}\right)\right]\\
&\approx \frac{\bar\sigma^2}m \frac{6\sqrt{3}}{4\pi}\left(\frac{T}m\right)^2e^{-2\frac{m}T} =\frac{\bar\sigma^2}\Delta\frac{3\sqrt{3}}{2\pi}\left(\frac{\kB T}\Delta\right)^2e^{-2\frac\Delta{\kB T}}\,.
\end{split}
\ee

The correction terms to this result are the third order $S_{23}+S_{32}$. However, these terms contain singularities for which the regularization has not yet been worked out explicitly. But as we discussed, unlike the case of the broadening of the Dirac-delta in the zero temperature DSF, there is no physical reason why unexpected singularities should show up in the higher terms, thus we stop at the second order.

\section{Exact transverse DSF in the field theory}
\label{app:FTexact}

Using the plane wave expansion Eq.~\eqref{psiexp} and $\eps=i\bar\psi\psi,$ the connected correlation function
\be
C^\eps(x,t)=\langle \eps(x,t) \eps(0,0)\rangle_T-\langle\eps(x,t)\rangle\langle\eps(0,0)\rangle_T
\ee
can be written as a four-fold rapidity integral of a linear combination of thermal expectation values of products of four creation/annihilation operators. Using the thermal Wick's theorem,
\be
\langle a^\pm(\th) b_1 b_2 b_3\rangle_T
= f^\pm(\th) \Big( \{a,b_1\}\langle b_2b_3\rangle_T - \{a,b_2\}\langle b_1b_3\rangle_T +
\{a,b_3\}\langle b_1b_2\rangle_T \Big)\,,
\ee
where $a^+=a^\dag,a^-=a,$ and $f^+(\th) =(1+e^{\beta\eps_\th})^{-1}= f(\th),$ $f^-(\th) =(1+e^{-\beta\eps_\th})^{-1}=1-f(\th),$ one arrives at
\be
\begin{split}
C^\eps(x,t)&=\frac{m^2}{4}\int\frac{\ud\th}{2\pi} \int\frac{\ud\th'}{2\pi}
\left[
f(\th)f(\th')(e^{\th-\th'}-1)e^{i(p+p')\cdot x}+
(1-f(\th))(1-f(\th'))(e^{\th-\th'}-1)e^{-i(p+p')\cdot x}\right.\\&
\left.+f(\th)(1-f(\th'))(e^{\th-\th'}+1)e^{i(p-p')\cdot x}+
(1-f(\th))f(\th')(e^{\th-\th'}+1)e^{-i(p-p')\cdot x}
\right]\,,
\end{split}
\ee
where we used the Lorentz product notation, $p\cdot x=\eps t - p x.$
At zero temperature $T=0,$ $f(\th)=0$ and we obtain the closed form result
\be
\begin{split}
C(x,t)&=\frac{m^2}{4}\int\frac{\ud\th}{2\pi} \int\frac{\ud\th'}{2\pi} (e^{\th-\th'}-1)e^{-i(p+p')\cdot x}
=\frac{m^2}{4}\int\frac{\ud\th}{2\pi} e^\th e^{-ip\cdot x} \int\frac{\ud\th'}{2\pi} e^{-\th} e^{-ip'\cdot x}-\frac{m^2}{4}\left(\int\frac{\ud\th}{2\pi} e^{-ip\cdot x}\right)^2\\&
=m^2K_0^2\left(m\sqrt{x^2-t^2}\right)-m^2K_1^2\left(m\sqrt{x^2-t^2}\right)\,.
\end{split}
\ee
Note that since $(p+p')\cdot x = (\eps_\th+\eps_{\th'})t-(p_\th+p_{\th'})x$ and $\eps_\th+\eps_{\th'}\ge2\Delta,$ after Fourier transformation $S(\w,q)|_{T=0}=0$ for $0<\w<2\Delta.$
At low temperature, the leading order can be obtained by approximating $f(\th)\approx e^{-\beta\eps(\th)}$ and keeping only first powers of $f(\th):$
\be
\begin{split}
C(x,t) & \approx \frac{m^2}{4}\int\frac{\ud\th}{2\pi} \int\frac{\ud\th'}{2\pi}
\left[
(1-f(\th)-f(\th'))(e^{\th-\th'}-1)e^{-i(p+p')\cdot x}\right.\\&
\left.+f(\th)(e^{\th-\th'}+1)e^{i(p-p')\cdot x}+
f(\th')(e^{\th-\th'}+1)e^{-i(p-p')x}
\right]\,.
\end{split}
\ee
Taking the Fourier transform, for frequencies $|\omega|<m$ only the second line gives nonzero contribution and it recovers the expression Eq.~\eqref{C11}.

\section{Asymptotic analysis of the integral Eq.~\texorpdfstring{\erf{sxx}}{}}
\label{app:asympt}

In this appendix we report the details of the asymptotic analysis of Eq.~\erf{sxx} for the local transverse DSF. We approximate the integral by dividing it into two integrals at the extreme point
$q_m=\sqrt{\frac{\omega\Delta }{J (\Delta +2 J)}}$ of the exponent:
\be
\begin{split}
S^{xx}(\omega)&\approx
\int_{q_c}^{\pi}\frac{d q}{\pi}\
\left| \frac{\sqrt{1 + g^2-2 g \cos \frac{q}{2}}}{ g J \sin \frac{q}{2}} \right|
\,e^{-\epsilon_{k_-+q/2}/(\kB T)}
\\&\approx\int_{q_c}^{\pi}\frac{d q}{\pi}\
\frac{\frac{\Delta }{J q} + q \left(\frac{J}{2\Delta }+\frac{\Delta }{24J}+\frac{1}{2}\right) }{\frac{\Delta }{2}+J}
\cdot \exp\left\{-\frac{1}{\kB T}  \left[\Delta +\frac{\Delta}{4 J (\Delta +2 J)}\frac{\omega^2}{q^2} + q^2 J\left(\frac{J}{2\Delta }+\frac{1}{4}\right)\right]\right\}
\\&\approx\int_{q_c}^{q_m}\frac{d q}{\pi}\
\frac{\frac{\Delta }{J q} + q \left(\frac{J}{2\Delta }+\frac{\Delta }{24J}+\frac{1}{2}\right) }{\frac{\Delta }{2}+J}
\cdot \exp\left\{-\frac{1}{\kB T}  \left[\Delta +\frac{\Delta}{4 J (\Delta +2 J)}\frac{\omega ^2}{q^2} \right]\right\}
\\&+
\int_{q_m}^{\pi}\frac{d q}{\pi}\
\frac{\frac{\Delta }{J q} + q \left(\frac{J}{2\Delta }+\frac{\Delta }{24J}+\frac{1}{2}\right) }{\frac{\Delta }{2}+J}
\cdot \exp\left\{-\frac{1}{\kB T}  \left[\Delta + q^2 J\left(\frac{J}{2\Delta }+\frac{1}{4}\right)\right]\right\}
\\&=
\frac{1}{\pi}\frac{e^{-\frac{1}{\kB T}  \left(\Delta +\frac{\Delta  J}{\Delta +2 J}+\frac{\omega }{4}\right)}}{96 \Delta  J^3 (\Delta +2 J)^2}
\left[ \omega  \left(\Delta ^2+12 J^2+6 \Delta  J\right) \left(4 \Delta  J e^{\frac{ \Delta J}{\kB T(\Delta +2 J)}}-\omega  e^{\frac{ \omega }{4\kB T}} (\Delta +2 J)\right)
\right.\\&\left.
-\Delta  J e^{\frac{ \omega }{4\kB T}+\frac{ \Delta  J}{\kB T(\Delta +2 J)}} \left(96 \Delta  J (\Delta +2 J)-\frac{\omega ^2}{\kB T}  \left(\Delta ^2+12 J^2+6 \Delta  J\right)\right) \left(\Gamma \left(0,\frac{J \Delta }{\kB T(2 J+\Delta) }\right)-\Gamma \left(0,\frac{\omega }{4T}\right)\right) \right]
\\&+
\frac{1}{\pi}
\frac{e^{-\frac{1}{\kB T}  \left(\Delta +\frac{\pi ^2 J (\Delta +2 J)}{4\Delta }+\frac{\omega}{4} \right)}}{6  J^2 (\Delta +2 J)^2}
\left[6  \Delta  J (\Delta +2 J) e^{\frac{1}{\kB T}  \left(\frac{\pi ^2 J (\Delta +2 J)}{4\Delta }+\frac{\omega}{4} \right)} \left(\text{Ei}\left(-\frac{J \pi ^2  (2 J+\Delta )}{4 \Delta T }\right)-\text{Ei}\left(-\frac{\omega}{4\kB T}\right)\right)
\right.\\&\left.
+\kB T\left(\Delta ^2+12 J^2+6 \Delta  J\right) \left(e^{\frac{\pi ^2  J (\Delta +2 J)}{4 \Delta \kB T }}-e^{\frac{ \omega }{4\kB T}}\right) \right]
\\&\approx
\frac{1}{\pi}e^{
-\frac{\Delta}{\kB T} }
\left(
-\frac{2 \Delta}{J (\Delta +2 J)}\left(\ln \left(\frac{\omega}{4\kB T} \right)+\gamma_E \right)+
\frac{\Delta ^2+12 J^2+6 \Delta  J}{6   J^2 (\Delta +2 J)^2}\kB T
\right)+\cdots
\\&\approx
\frac{\Delta}{\pi J^2}
e^{-\frac{\Delta}{\kB T} }
\left[ -\ln\left(\frac{\omega}{4\kB T}\right)+\frac{\kB T}{2\Delta}-\gamma_E \right]+\cdots
\,,\end{split}
\ee
with the
incomplete gamma function $\Gamma (a,z)=\int _z^{\infty }d t\ t^{a-1} e^{-t}$
and the exponential integral function $\text{Ei} (z)=-\int_{-z}^{\infty } \frac{e^{-t}}{t} \, dt$.
The last line is obtained by taking scaling limit $\Delta/J\rightarrow0$,
namely, $g \to g_c$, and the result agrees with field
theory result Eq.\eqref{C11asym}.

\section{NMR relaxation rates for large nuclear spin}
\label{app:NMR}

For nuclear spin $I > 1/2$, the nuclear quadrupole interaction splits the nuclear spin energy levels, and Eq.(\ref{eq:T2rate})
needs
 to be
evaluated
based on Bloch-Wangsness-Redfield theory using the density matrix for nuclear spin $\rho_{\alpha\alpha^{\prime}}$ \cite{SlichterBook,PenningtonPRB1989,PenningtonThesis,BarretThesis},
\begin{equation}
\frac{d\rho_{\alpha\alpha^{\prime}}}{dt}=\sum\limits_{\beta\beta^{\prime}}^{}R_{\alpha\alpha^{\prime},\beta\beta^{\prime}}\rho_{\beta\beta^{\prime}},
\label{eq:Redfield1}
\end{equation}
where $\alpha$, $\alpha'$, $\beta$ and $\beta'$ specify the nuclear spin energy levels, and $R_{\alpha\alpha^{\prime},\beta\beta^{\prime}}$ is the element of the relaxation matrix $R$.  In this approach,
$
1/T_{2}
 = R_{\alpha \alpha-1, \alpha  \alpha-1}$ for the $I_{z}=\alpha$ to $\alpha-1$ transition of a given nuclear spin $I$ \cite{SlichterBook}, and
\begin{equation}
\frac{1}{T_{2}
}
= A~\frac{1}{T_{1}
}
+ \gamma_{n}^{2}~\overline{h
^{2}}~\tau_{o},
\label{eq:Redfield2}
\end{equation}
where the pre-factor $A$ is a constant that depends on $I$,
$\gamma_n$ is the nuclear gyromagnetic ratio of the observed nuclear spin,
$\overline{h^{2}}$
 represents the averaged fluctuating hyperfine magnetic field along the
direction of the external magnetic field (i.e. $x$-axis in the present case of TFIC),
$\tau_\text{o}$ is the correlation time ($\omega_{n}\tau_\text{o} \ll 1$), and the second term represents $1/T_{2}^{\prime}$ within
the framework of
Redfield's theory.
In the case of nuclear spin $I=1/2$ with no nuclear quadrupole splitting, $A=1/2$ \cite{SlichterBook}.
For the $I_{z}=+1/2$ to  $-1/2$ central transition of $I=3/2$, earlier work showed that $A=7/2$ \cite{PenningtonPRB1989,PenningtonThesis}.
In the case of $I=9/2$ at $^{93}$Nb sites in the TFIC candidate material CoNb$_2$O$_6$,
the calculations of
$A$
 are straightforward but rather tedious, and we obtained $A=49/2$.

\bibstyle{apsrev-nourl}
\bibliography{bib_transverse}

\end{document}